# Business models to assure availability of advanced superconductors for the accelerator sector and promote stewardship of superconducting magnet technology for the US economy

A report sponsored by the US Department of Energy
Office of Accelerator Research Development and Production

April 2023

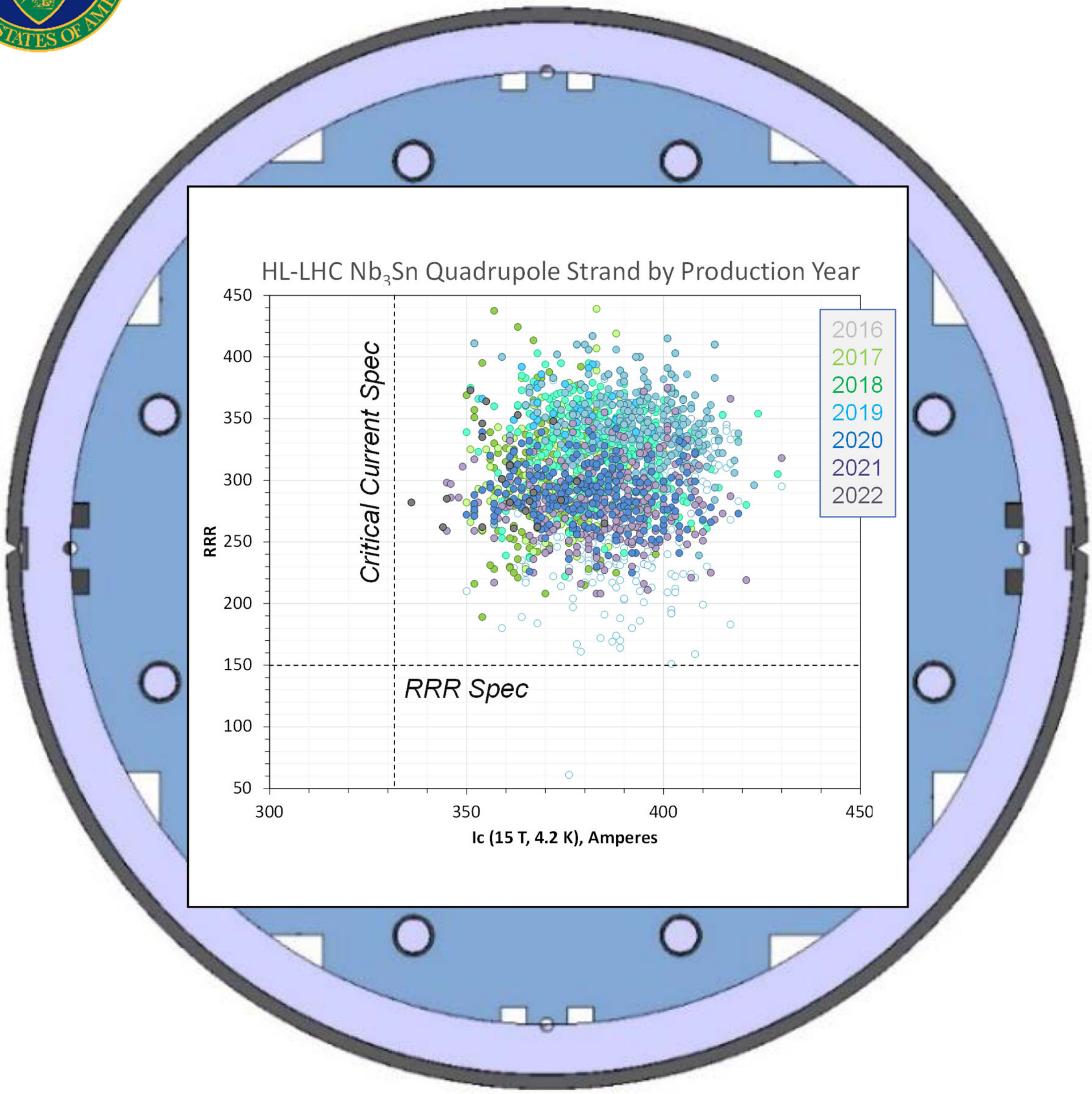

*This study is based on work supported by the Department of Energy Office of Science under Grant No. DE-SC0022298.*

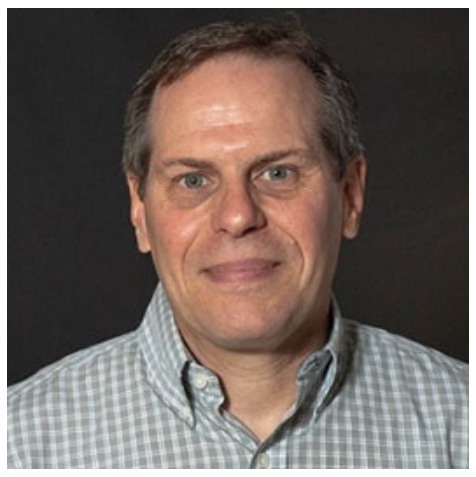

Lance Cooley
Professor – Florida A&M University and Florida State University
College of Engineering
Director – Applied Superconductivity Center
National High Magnetic Field Laboratory
Tallahassee, FL 32310

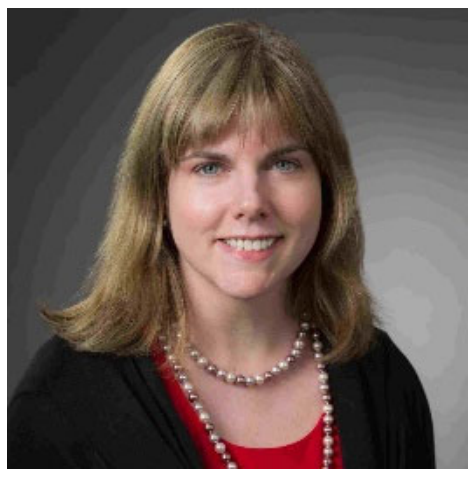

Kathleen Amm
Director – Superconducting Magnet Division
Brookhaven National Laboratory
Upton, NY 11973

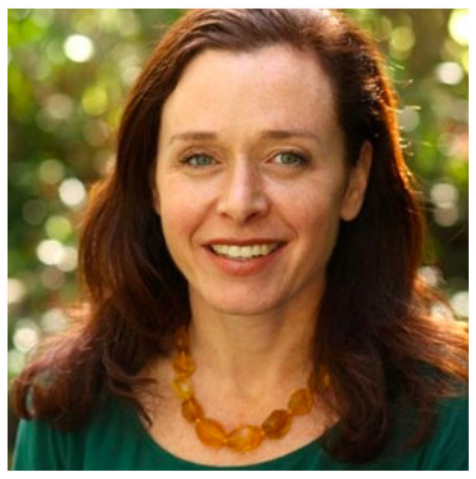

Whitney Hischier
Lecturer – Haas School of Business
University of California, Berkeley
Berkeley, CA 94720

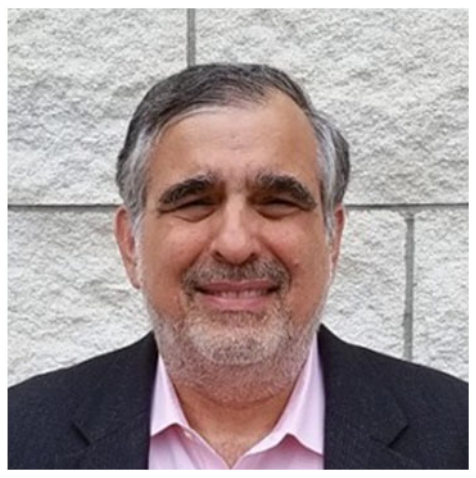

Steven Rotkoff
Colonel – United States Army (Ret.)
Red Teaming by Rotkoff

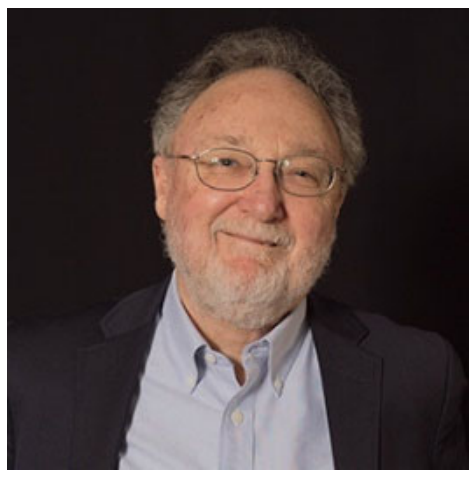

David Larbalestier
Professor – Florida A&M University and Florida State University
College of Engineering
Chief Materials Scientist – National High Magnetic Field Laboratory
Tallahassee, FL 32310

(this page left blank intentionally)

# Contents

(this page left blank intentionally)

## Executive Summary

***Stakeholders representing concerns of national and global leadership, industries that use superconducting magnets in products, manufacturers of superconducting wires and tapes that supply to industries, and innovation generators from small businesses and universities came together to address stewardship of superconducting magnet technology and assurance of supply of advanced superconductors to the accelerator sector. This report outlines potential public-private partnerships that develop and enhance domestic capabilities to meet the needs of science facilities in the accelerator systems sector and in the broader commercial ecosystem.***

***Superconducting magnet technology and its related conductor technology are strategic and critical technologies for global leadership. In addition to science, technologies that use magnets are common in the energy, transportation, defense, manufacturing, health and medicine, and space sectors. Investment in science facilities creates an economic return 5-20 times the investment in the facility. This excludes additional impact produced by facilitating the emergence of industries like medical imaging magnets. Business sectors in energy and medicine are at tipping points for rapid economic growth, with >$1 billion private investment in research.***

***Magnet and conductor development programs under the US Department of Energy Office of High Energy Physics have succeeded in advancing $Nb_3Sn$ technologies from concept to production of tunnel-ready 11 T magnets within 15 years using sustained investments of ~$12 million per year. While this model defines the present business plan for continued stewardship of superconducting magnet technology, opportunities in particle physics requiring 16 T and 20 T magnets motivate expansion of the program scope to add three future technology options. At the same time, flat funding and increased costs for high-temperature superconductor options greatly constrict the ability of the present plan to succeed. Four scenarios were envisioned by the stakeholders:***

1. ***A status quo scenario could result in technology stagnation and atrophy of the present innovation cycle. Market factors could cause the primary conductor supplier to the accelerator sector to consolidate conductor manufacturing outside of the scope of federal conductor R&D support, thereby weakening the innovation cycle. A gap of 5 or more years is foreseen before alternative supply chains can be developed, without assurance that present 12 T technology can be re-attained.***
2. ***A scope reduction scenario could result in limiting options for advanced conductors and increasing cost, risk, and uncertainty for future magnet technologies. Advancement of technology to be ready for future facilities could require significant re-development of existing roadmaps.***
3. ***A scenario with expansion of public funds to $30 million annually could be sufficient to map the prior HEP success across parallel programs toward industry-built 12 T and laboratory-built 16 T and optional 20 T magnets. Increased stewardship of conductor development could have important tipping factors for emergence of technologies in multiple areas. Following this scenario is recommended by stakeholders.***
4. ***A scenario examined the present billion-dollar private investment in REBCO conductors for fusion. If sustained, this investment threatens to displace present markets that presently rely upon the same conductors used in accelerator magnets. Follow-on investments to support other energy sector applications, avoidance of liquid helium supply chain risks, and avoidance of reaction furnaces for magnets, could close off the industrial ecosystem that the accelerator sector relies upon. Investment of $10 million annually could allow the accelerator sector to engage with research into helium-free magnets and magnets for a potential muon collider, which overlap more strongly with commercial magnets. A combination of scenarios 3 and 4 above places the United States in the strongest competitive position for science facilities and technology advancement.***

***A new business plan could operate with minor modifications of the existing HEP programs that facilitate public-private partnerships. The present framework has effective management, intellectual property development methods, inventory control, specifications, quality plans, testing infrastructure and expertise, strong couplings with university groups, and administrative support from national laboratories. Stakeholders requested augmentation of the framework to include:***

- ***Bringing industry personnel into national laboratory projects, and vice-versa, to facilitate stewardship of the industry ecosystem.***
- ***Improving industry access to national laboratory and university infrastructure, possibly under existing frameworks for partnerships and service agreements.***
- ***Retraining of industry personnel at universities that traditionally pipeline talent and increasing engagement of future workforce with industry.***
- ***Consideration of manufacturing centers of excellence or innovation institutes connected with universities and laboratories***
- ***Improve marketing and communication to attract young talent to the superconductivity marketplace and the accelerator sector.***

## Importance of the accelerator sector to the development of superconducting magnet technology

The accelerator sector creates technologies to accelerate particle beams for scientific and industrial purposes. It is a leading developer of superconducting magnet technology because of the need to steer, focus, and otherwise constrain charged particle beams up to the inconceivably high energies that existed at the birth of the universe. Some accelerators are among the largest and most complex machines conceived and built by humankind, with technology that impacts industry paradigms for decades after the construction of a major facility. Large accelerator facilities must address the needs of physicists conducting research at the frontiers of knowledge, which drives both the operation of superconducting magnets at the limits of technology as well as the continual development of magnets with ever greater capability. The accelerator sector therefore pioneers magnet and associated conductor technology using roadmaps a decade or more ahead of envisioned applications in industry. Public investments in the accelerator sector can thus catalyze the emergence of commercial markets and accelerate their maturation due to the impracticality of sustaining similar roadmaps over long terms in industry. In return, projects in the accelerator sector benefit from manufacturing scale and cost reduction driven by commercial economics.

Magnets, the materials to make magnets, and technologies that use magnets are of strategic importance to the United States [1]. Technologies that use magnets are common in the energy, transportation, defense, manufacturing, health and medicine, space, and science and technology sectors. The economic value of permanent magnets alone was over $19 billion in 2021 and is expected to have a compound annual growth rate (CAGR) of over 7% between 2022 and 2030 [2]. Superconducting magnets (hereafter "magnets") define a unique part of magnet technology because no other technology can provide access to high magnetic fields for extended duration. "High field" (HF) identifies magnetic field strength above the saturation magnetization of iron and permanent magnet materials, i.e. above 1.5 T field. High-field magnets are electromagnets wound from composite wires and tapes that contain superconducting materials (see Appendix 1), where the field strength and physical size of the magnet determines the amount of conductor needed. "Ultra-high field" (UHF) applies to applications above 20 T, which is beyond the technology limit for the established industrial wire superconductors (herein "magnet conductors" or "conductors") using Nb-Ti alloy or the compound $Nb_3Sn$. Ultra-high field magnets require the use of high-temperature superconductors (HTS) because they retain superconductivity up to 100 T or higher when operated at low temperature, e.g. when cooled by liquid helium to 4.2 K. Magnets made from HTS windings are also capable of high fields when operated at higher temperature, e.g. 20 T at 20 K and 6 T at 50 K. Appendix 1 summarizes the conductors in the present supply chain and their stages of readiness [3].

The accelerator sector relies upon high-field magnets, presently at 12 T, and envisions high field magnets at 16 T and ultra-high field magnets at 20 T or beyond. Large accelerators require thousands of magnets to steer and focus particle beams and thousand-ton procurement runs to supply conductors. An envisioned Future Circular Collider (FCC), for instance, would have a circumference of approximately 100 km and require ~ 5,000 dipole magnets each being 15 m long, where every meter of magnet would require 115-150 kg of $Nb_3Sn$ conductor [4]. Another ~750 quadrupoles would be needed. The conductor quantity for the dipoles alone would be above 10,000 metric tons based on these values. At the present cost of ~$2 million per ton of $Nb_3Sn$ conductor, the FCC represents a potential $20 billion market for conductor alone. Even a small fraction of this procurement would have a transformative effect on the manufacturing of $Nb_3Sn$ and HTS wires and tapes.

Moreover, the accelerator sector is one pillar of a broader nexus of activities that serve researchers in fundamental science by constructing large facilities to address grand challenges. The second pillar, fusion, has driven magnet and conductor technology roadmaps to build tokamak reactors. The International Thermonuclear Experimental Reactor, ITER, drove a 600-ton conductor procurement of low-loss $Nb_3Sn$ superconducting wire (requirements are different than accelerators), and a future demonstrator for power production could require 10 times this amount [5]. Private resources are investing ~ $2 billion in the possibility of very compact ultra-high field tokamaks, which is presently driving the HTS conductor market to increase scale by a factor of 10. The third pillar is ultra-high field science, which includes exploration of novel materials, biology, chemistry, and quantum matter at dc fields presently up to 45 T and in pulsed fields above 100 T [6]. Superconducting solenoid magnets capable of 40 T are presently being researched.

To advance technology for envisioned physics needs, the accelerator sector intentionally drives technology advancement through basic and directed R&D programs that link national laboratories, universities, and industry. This is discussed in more detail in Appendix 1, and the effect is shown schematically in Figure 1. Advanced technology development has been carried out for at least 4 decades under the guidance of the US Department of Energy (DOE) Office of High Energy Physics (HEP), using community workshops such as the Low Temperature Superconductor Workshop (LTSW), the Workshop on Mechanical & Electromagnetic Properties of Composite Superconductors (MEM), and programs such as the Magnet Development Program (MDP) and the Conductor Procurement and R&D program (CPRD) to nurture technology advancement. The other pillars supporting superconducting magnet development have occasionally networked with HEP programs, so building more coherence across these activities for superconducting magnet technology development could lead to faster rates of advancement. Nonetheless, the HEP R&D effort applied several successful principles and achieved outcomes sufficient to build the High-Luminosity upgrade of the Large Hadron Collider (LHC), called "Hi-Lumi" or "HL-LHC". The $250 million US contribution, called the HL-LHC Accelerator Upgrade Project or "AUP", is midway to completion, with the 10-ton conductor procurement completed and the cable fabrication 95% completed (as of Dec. 2022). In summary:

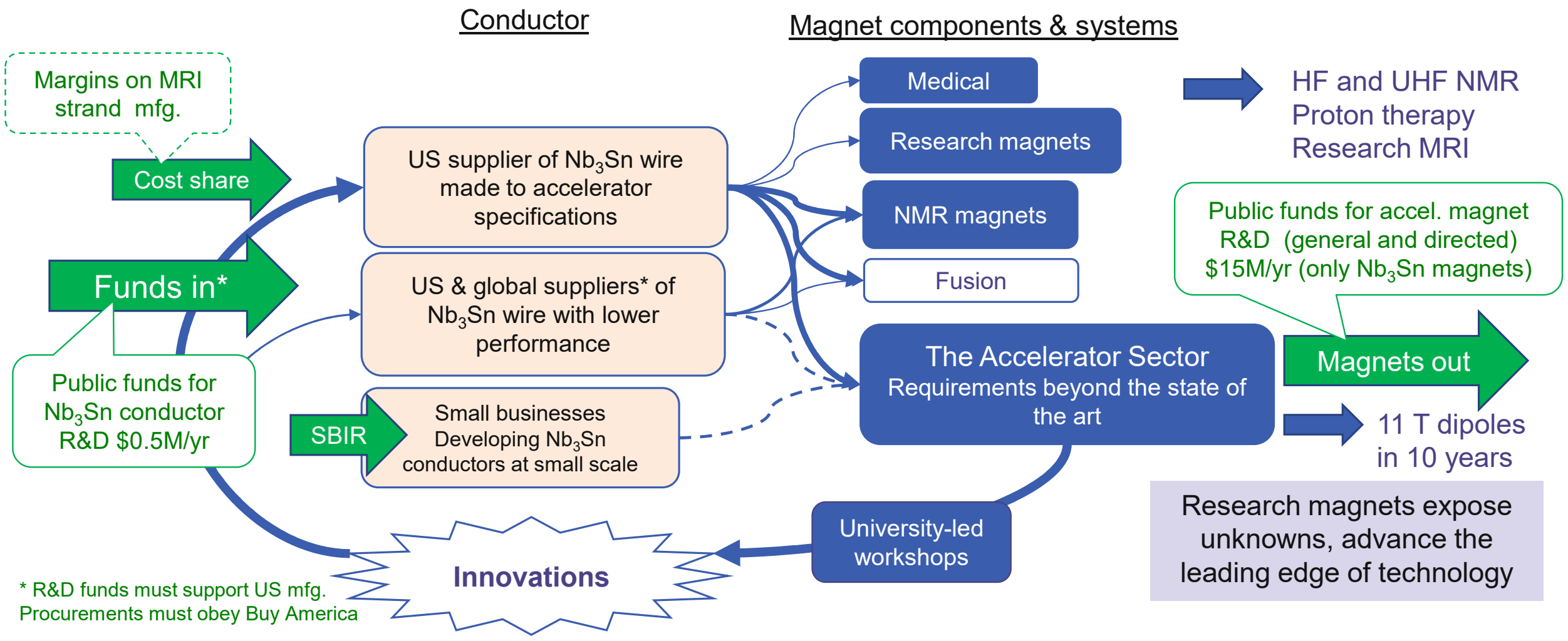

**Figure 1:** This virtuous cycle of $Nb_3Sn$ conductor and magnet technology development existed during the period leading up to the High-Luminosity Upgrade (Hi-Lumi or HL) of the Large Hadron Collider (LHC). Investment of $4 million in conductor R&D between 2003 and 2014 led to the development of accelerator grade $Nb_3Sn$ conductor and enabled the pre-production scale-up during 2015–2016 and the full production run between 2017 and 2022. While multiple manufacturers were initially considered, only one manufacturer met performance targets early on. The conductor architecture evolved several times to address changing requirements of magnets as they also matured. The final conductor supports the present 11 T magnet technology being installed at CERN. Availability of high-performance conductor also supported the emergence of high-field and ultra-high field NMR systems now being sold by Bruker Biospin and medical therapy systems now being marketed by Mevion and Varian.

Important elements of the HEP development model:

- Basic and directed R&D based upon only 1 conductor type: accelerator grade $Nb_3Sn$.
- Sustained R&D funds for magnets at ~$12 million per year across the accelerator sector, including directed R&D funds from the LHC Accelerator Research Project, or LARP, at $6-7 million per year [7], [8].
- At least one significant magnet milestone achieved per year. Most milestones resulted in modifications in conductor specifications and testing that pulled development.
- Sustained R&D funds for conductors administered by the predecessor of CPRD called the Conductor Development Program or CDP. CDP furthermore coordinated effort with procurements by LARP. Overall, $4 million was spent on conductor development between 2005 and 2014 [9], followed by $4.5 million during 2015-2017 for pre-production and $16.8 million for full production [10].
- Cost-share by the conductor manufacturers, although by 2010 margins on MRI wire fabrication had diminished by market forces as those conductors entered the commodity stage of the product life cycle..
- Continuity of the workshop-innovation feedback cycle.
- Cross fertilization of the feedback cycle into program and project reviews.

Important achievements:

- Advancement of conductor technology from concept to procurement in 15 years [11].
- Advancement of magnet technology from concept to procurement in 15 years [12].
- Advancing the operating field strength of accelerator magnets from 8.6 T to 11.0 T [11].
- Highly reproducible manufacturing with process capability indices above 1.25. A process capability index is generated from manufacturing statistics, and a score above 1.25 indicates high yield and high efficiency suitable for a major acquisition in the accelerator sector.

What was not achieved:

- Emergence of large markets using accelerator-grade $Nb_3Sn$. Conductor demand for NMR magnets, laboratory science magnets, and medical therapy is comparable to the scale of the HL-LHC procurement.
- Cost reduction of the $Nb_3Sn$ product.

## The need for a new business plan: An innovation cycle under stress

The United States has led the global development of advanced accelerator, fusion, and high-field science magnets for several decades. This leadership has directly contributed to leadership in medical imaging and therapy, biology, chemistry, genomics and pharmaceuticals, petroleum, electricity, defense, transportation, space, and science. However, **investment in advanced magnet development has become stagnant at a time when expansion in scope is motivated by physics challenges** [11],[15]. Flat funding for superconducting magnet technology R&D in the accelerator sector is reflected in Figure 2, where superconducting magnet R&D (SC Magnet) is about 1/5 of the overall budget allocated to General Accelerator R&D (GARD) [7]. Funding for university programs working on superconducting magnets and conductors, not shown in Fig. 2, has also been flat at ~$5 million per year. At the same time, physicists contemplate needs for 16 T magnets and options for 20 T [4] or higher magnets (e.g. 50 T magnets in a muon collider [16], [17]), which requires development of technology alternatives using advanced $Nb_3Sn$ wire [13] and two HTS conductor types. The accelerator grade of $Nb_3Sn$ wire is mature so increasing its

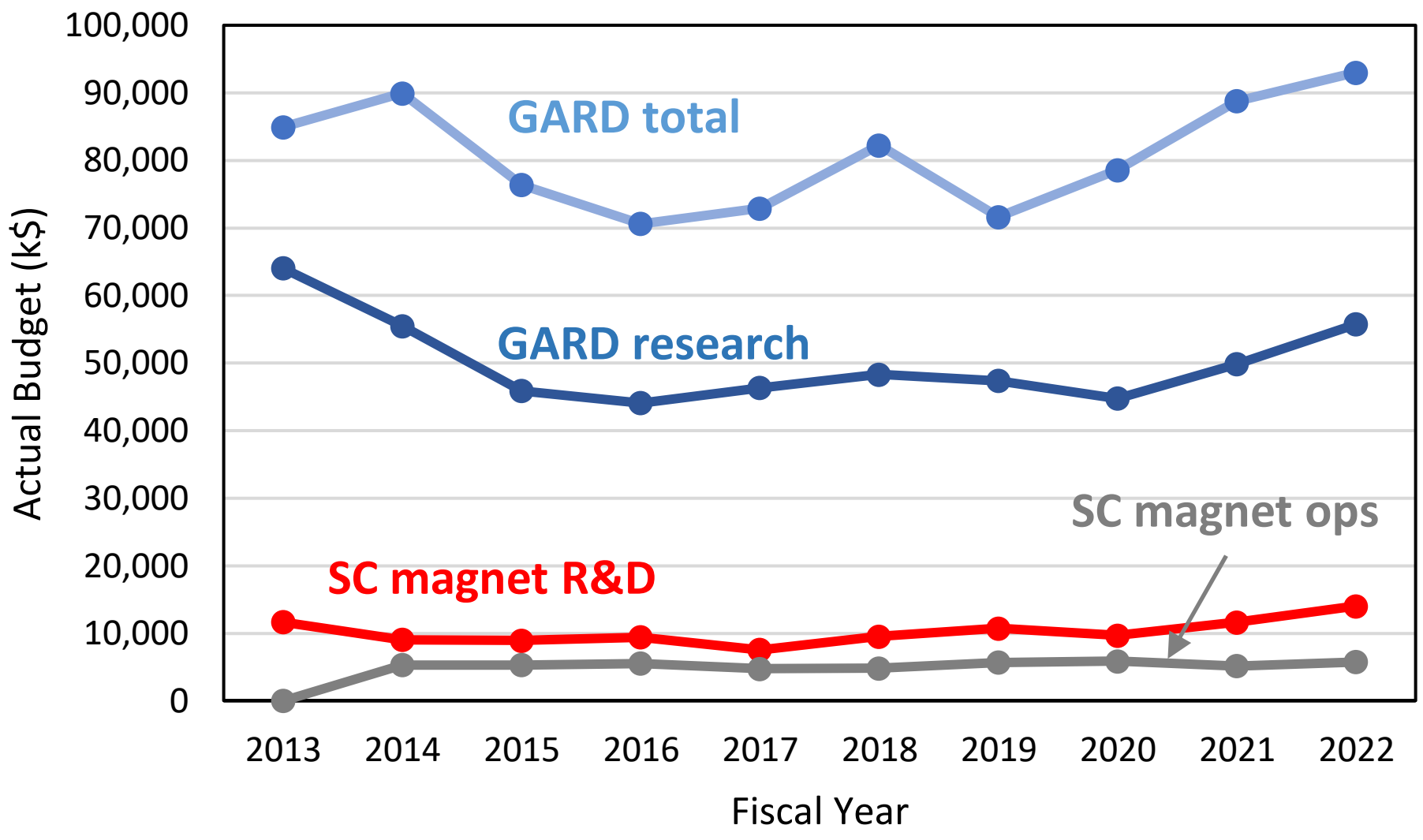

**Figure 2:** Funding profile (in-year dollars as appearing in federal budget documents) for General Accelerator R&D (GARD) supplied by the HEP office for this report. The portion allocated to superconducting magnet and conductor R&D is also indicated in gray. The allocation to magnet operations supports test facilities such as those at Fermilab.

performance further by ~30% in an advanced grade of conductor will be difficult. Also, HTS conductors are expensive, five times the cost of $Nb_3Sn$ (see Appendix 1), so these additions have significant impact even at small programmatic scale.

The innovation cycle is under additional stress due to inflation. Costs for labor, raw materials, operations, and maintenance have risen by over 50% since the time LARP began. The present cost of 1 kiloampere-meter of accelerator grade $Nb_3Sn$ wire for a 12 T operating field now exceeds $13, up from a low point of $7.50 in 2005 with an annual escalation rate of over 3%. This is represented in Fig. 3. Prior to the achievement of 3,000 A mm$^{-2}$ current density in the non-copper area in 2003, the cost per kA-m was steeply declining due more to the increase of performance (more kilo-amperes for the same cross-section) than reduction of the cost per meter. Lack of large-market demand to drive increased production scale has prevented significant reduction of cost per meter. Dilution of funds for innovation across multiple conductor types requested by the accelerator sector has added further stress. This is compounded by the fact that HTS conductors have a cost per meter 5 times (or more) higher than that of $Nb_3Sn$ conductors.

Stakeholder meetings, see Appendix 2, were held in early 2022 to receive input about possible business plans to address national security and global leadership, magnet industry growth, conductor manufacturing, and innovation and research. Scenarios developed with the assistance of stakeholders are presented below to illustrate the potential outcomes of the present development plan as well as outcomes for alternative public-private partnerships for the accelerator sector.

## *Scenario 1 – Status Quo*

This scenario is illustrated in Figure 4. Flat funding continues at the same level as present, with no change to address dilution across multiple magnet technology options or the increased cost of HTS options. **The present outlook by the stakeholder community corresponds to Figure 4, which leads to undesirable outcomes.**

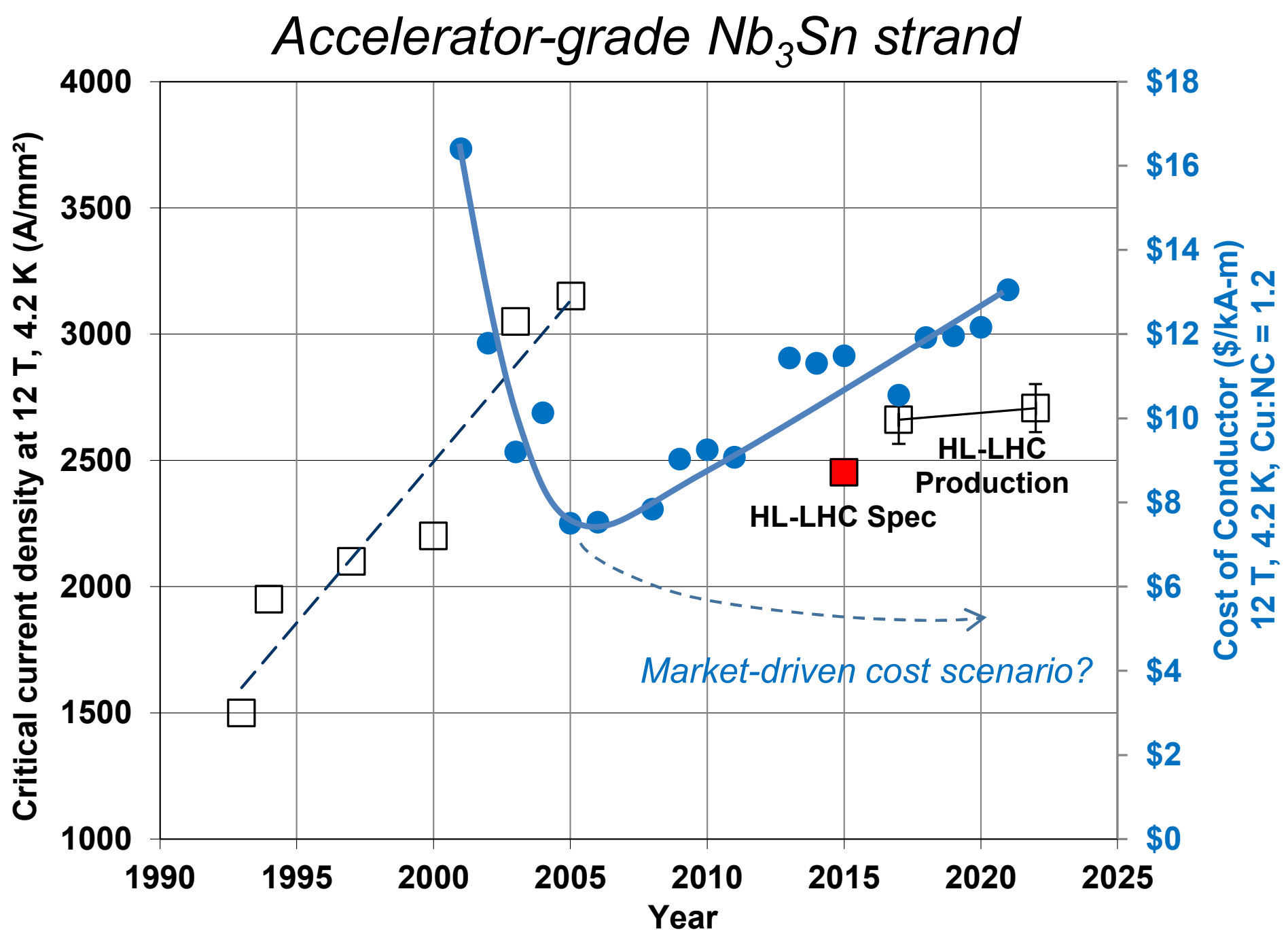

**Figure 3:** Data for champion critical current density achievements and HL-LHC specifications are shown as square symbols referenced to the left y-axis. The cost per kiloampere-meter drawn from national laboratory procurement records for accelerator-grade $Nb_3Sn$ strand is also shown as blue circles referenced to the right y-axis. During 1994-2004, CDP-supported innovations resulted in a significant increase in performance and a concomitant reduction of cost per kA-m. However, despite the large quantity of the HL-LHC strand procurement, absence of large markets demanding $Nb_3Sn$ strand has not

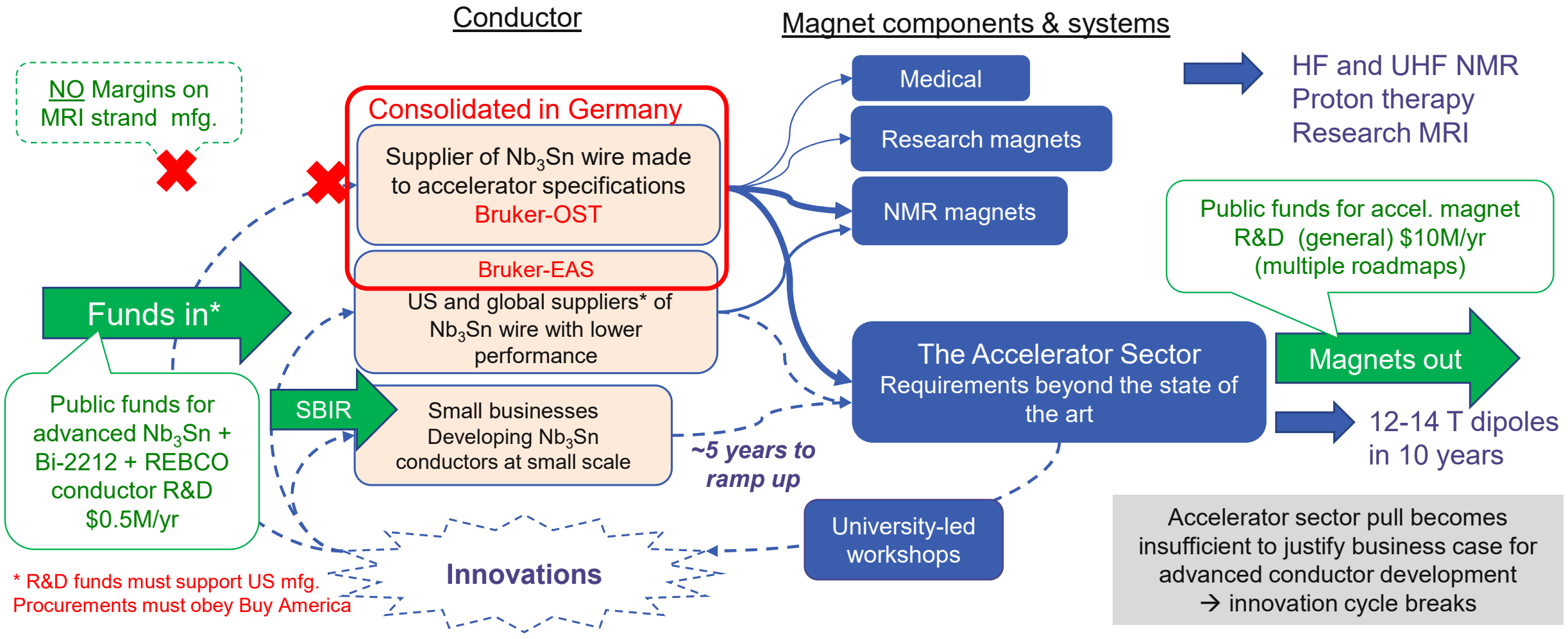

**Figure 4:** Keeping the status quo, investment in accelerator sector magnet and conductor R&D programs continues at levels established in 2015, which become diluted across multiple development roadmaps and challenged by high cost of HTS conductors. If economic conditions evolve such that Bruker consolidates manufacturing into a single factory outside of the domain where US public funds can support development, then a key developmental pathway will be severed.

Actions required: None

Advantages:

1. Easiest option. Assumes present US federal budget remains stable or increasing.
2. All conductor options continue present roadmaps.
3. Magnet options to meet physics targets continue present roadmaps.

Disadvantages:

1. Funds are diluted below critical levels of investment for some or all conductor types, potentially leading to:
    a. Atrophy of the innovation loop.
    b. Lack of manufacturing innovation to meet emerging market needs.
2. Potential erosion of the business case to justify shoulder products for the accelerator sector. Manufacturing of MRI wire and support of NMR market dominate business decision influence.
    a. Companies consolidate to support MRI and NMR businesses. The primary supplier of $Nb_3Sn$ conductor, Bruker OST in Carteret, NJ, is consolidated to overseas manufacturing as soon as 2024.
    b. Since Bruker OST supplied all the HL-LHC $Nb_3Sn$ wire and all the advanced $Nb_3Sn$ and Bi-2212[i] research wire to HEP programs, conductor development roadmaps could be interrupted.
    c. A technology gap of ~5 years opens before alternative manufacturing can emerge.
3. Magnet and conductor technology stagnates in the accelerator sector.
    a. Target of 16 T $Nb_3Sn$ dipole delayed past 2030.
    b. Potential 20+ T HTS magnets must rely upon the very difficult REBCO[ii] conductor.
    c. Insufficient technology advancement for some future collider options desired by physicists.
4. Technology sharing with other sectors, e.g. NMR magnets, may be reduced since their supply chains also become constricted.
5. Higher probability of reduced manufacturing capacity to supply conductor for a future large accelerator facility.
6. Cost of conductor remains high.

### *Scenario 2 – Reduction of scope*

In this scenario, flat funding continues, and the dilution of funds across multiple conductor types is addressed by removing the advanced $Nb_3Sn$ alternative. A scenario where reduction of scope by removing HTS is not considered in view of the multiple national panel reports [6], [17], [19] calling for its prioritization.

Actions required: Re-prioritization of MDP and other programs. Research into $Nb_3Sn$ magnets shifts toward perfecting and industrializing present ~12 T $Nb_3Sn$ technology. HTS is developed for 16 T and higher.

Advantages:

---

[i] A HTS conductor based on the superconductor $Bi_2Sr_2CaCu_2O_{10}$, see Appendix 1

[ii] A HTS conductor based on the superconductor $RE_1Ba_2Cu_3O_7$, where RE denotes the element yttrium, a rare-earth element, or blends thereof, see Appendix 1.

1. Reduction of scope makes more effective use of available funds. Program could be even more effective with reduction of scope and increase of funds.
2. Conductor R&D focuses on Bi-2212, and possibly REBCO cables.
3. Potential for stronger overlap with HTS magnet development programs for NMR magnets, fusion, and other applications.

Disadvantages:

1. High cost. Present HTS cost is 5 times that of $Nb_3Sn$ and 20 times that of MRI wire.
2. Low conductor technology readiness, and very low conductor manufacturing readiness.
3. Supply chain obstacles – low manufacturing readiness for HTS raw materials.
4. Very low magnet technology status. Single specimens of benchtop demonstrations exist.

Unknown:

1. Would Bruker OST spin off its present boutique manufacturing of Bi-2212 wire as a separate manufacturing entity?
2. If so, how would such an entity survive given that the annual market for the material is still in its infancy?

### Scenario 3 – Increased public investment

In this scenario, funding is increased to all replication of the innovation cycle reflected in Figure 1 *for each conductor technology*, advanced $Nb_3Sn$, Bi-2212, and REBCO. This scenario is illustrated in Figure 5. This scenario considers the possibility that magnet applications are at a tipping point where added investment facilitates bridging valley-of-death challenges.

Actions required:

1. Increase investment in accelerator sector magnet and conductor R&D programs to $30 million per year to support *each* roadmap element at levels comparable to those indicated in Figure 1. Investment plan should include accounting for inflation.
2. Investment in conductors at $4 million/year, including building a limited stockpile. The investment plan should include considerations for inflation.
3. Sustain and even grow university investment above $10 million/year to add technical training programs and facilitate expansion of present research programs.
4. Support manufacturing innovation centers with industry-laboratory-university mix to further support human capital development.

Advantages:

1. Viability of advanced $Nb_3Sn$ conductors, which extend present magnet technology from 12 T to 16 T, is determined within 2 years and paths to pre-production (see Appendix 1) are determined within 5 years. Orders for advanced $Nb_3Sn$ conductors are fulfilled on a regular basis, and the supply chain of advanced alloys is established.
2. $Nb_3Sn$ conductors supplied by multiple vendors are qualified for accelerator use, providing alternative designs and competitive pricing. The $Nb_3Sn$ supply chain is diversified.
3. Viability of Bi-2212 conductors for accelerator and lab magnets is determined within 2 years and scaled to pre-production within 5 years.
4. Bi-2212 strand manufacturing is secured for the long term as a viable business. The Bi-2212 supply chain is stabilized.

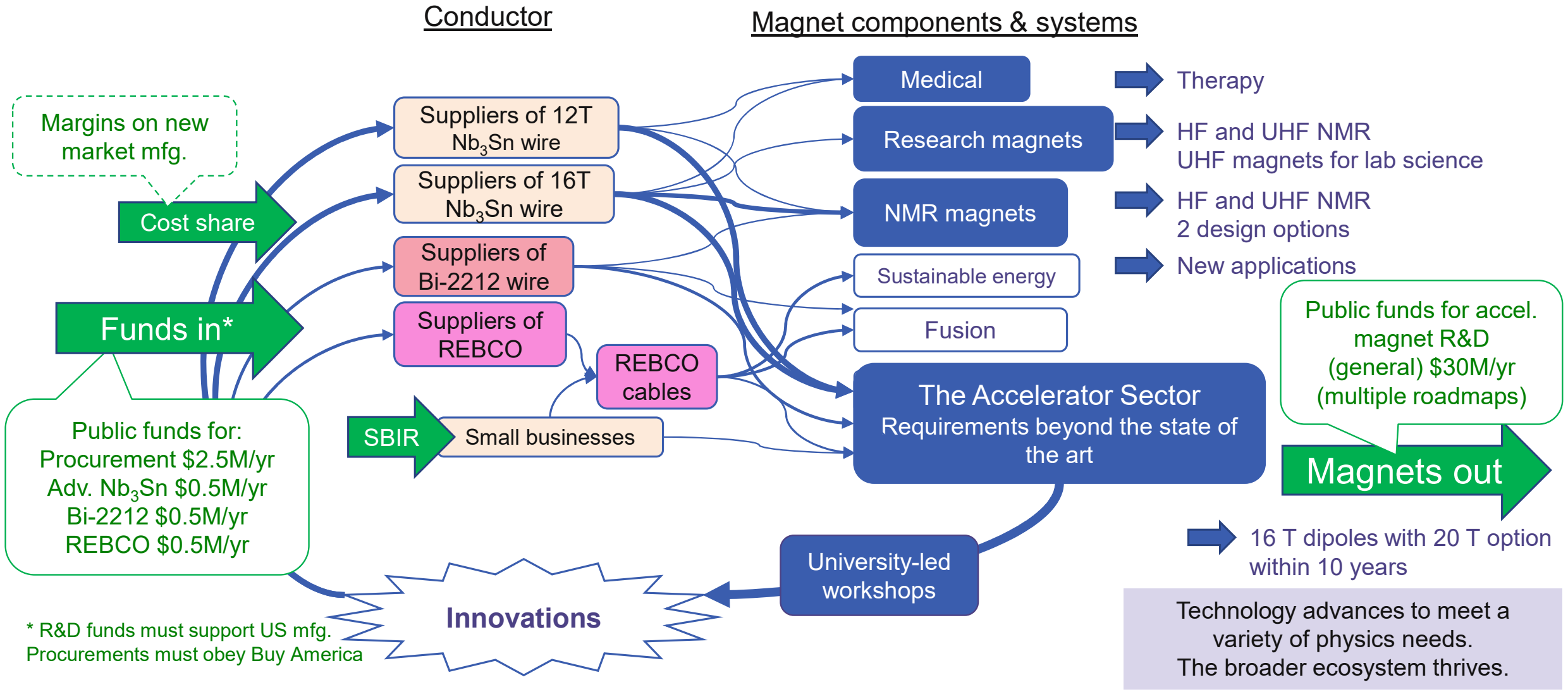

**Figure 5:** In Scenario 3, public investment in *each* accelerator sector magnet and conductor R&D roadmap is made at levels comparable to the 2002-2014 $Nb_3Sn$ activity, with appropriate further increases for inflation. The combined program grows to an estimated $30 million per year. Improved $Nb_3Sn$ conductors are matured, and conductors supplied by alternative vendors provide alternative designs, competition, and cost reduction. Options for ultra-high-field accelerators are provided by investment in Bi-2212 HTS conductors that are also manufactured by Bruker OST and SBIR companies. The high number of magnets constructed annually advances the technology at a pace sufficient to achieve physics goals in the 2040 timeframe. General advances across the accelerator sector-industry ecosystem also breed new markets as well as accelerate the deployment of UHF NMR, potentially creating a positive return on investment in economic growth.

5. Viability of present REBCO cable approaches for accelerator magnets is determined within 5 years, with significant benefit for fusion magnets.
6. A stockpile or national repository of conductor could be managed to circumvent supply chain issues for the accelerator sector and the industry ecosystem.
7. Magnet R&D achieves multiple significant milestones per year. Magnet technology advances at a faster pace potentially achieving physics goals in the 2040 timeframe.
8. General advances across the accelerator sector-industry ecosystem also breed new markets connected with liquid hydrogen and other applications in the 10-60 K range.
9. Advanced $Nb_3Sn$ and increasing maturity of ultra-high-field HTS conductors accelerate the deployment of UHF NMR, potentially crossing tipping points for chemical, pharmaceutical, and biological research.
10. Similar advances in medical therapy systems are more likely to be realized.
11. Partnerships between the national labs in the accelerator sector and commercial magnet industries are formed within 2 years, leading to faster advancement of emerging commercial applications in several sectors due to access to heavy assembly capabilities, test facilities, and expertise.
12. Conductor cost reduces via growth of markets that drive manufacturing at increased scale.

<u>Disadvantages</u>:

- Could require endorsements from advisory panel.

- Could require concurrence with Office of Management and Budget and Congress

### *Scenario 4 – Market disruption by private investment in REBCO.*

In this scenario, private investment in REBCO superconductors is sustained for at least one decade, leading to market disruption in favor of applications that take advantage of cost reduction and increase in REBCO manufacturing scale. While private investment in conductors and magnets for fusion is already at the scale indicated in Figure 6, a tipping point for other applications could be activated, leading to further private investment. Emerging applications include wind turbines and medical imaging magnets. The recent helium supply crisis further incentivizes technologies that do not rely upon a consistent helium supply. This could be very important for the accelerator sector, since a large future facility could require a significant fraction of the world's helium supply to be stockpiled [20]. Investment in hydrogen energy infrastructure could create strong interest in liquid hydrogen and 20 K applications, including emission-free aviation, ships, trains, trucks, and electricity microgrids. Since the accelerator sector presently relies upon magnet technologies using liquid helium, market disruption could constrain the conductor supply chain.

Actions required:

1. Coordinate the accelerator sector with the energy sector to ensure cross-fertilization of helium-free technology. The Office of Fusion Energy Sciences and the Office of Energy Efficiency and Renewable Energy are well positioned to facilitate coordination with the accelerator sector.
2. Mitigate risks of helium supply chain disruption and market constraints by expanding R&D program to encompass helium-free accelerator magnets, roughly $100 million over 10 years

Advantages:

1. Engagement with the energy sector leverages resources to transform and improves the REBCO conductor manufacturing maturity.
2. Meets the targets for future physics facilities while mitigating risks for long-term viability.
3. Avoids dependence on helium.
4. Provides a diverse supply chain at low cost.

Disadvantages:

1. High initial cost, with costs over decade time scales depending strongly on ability of private investments to drive cost reduction.
2. If private investment in REBCO is not sustained, addition of public funds indicated above could become a sub-optimal investment.
3. Compact fusion is high risk, high reward. Many technologies need to be invented to realize it including multiple advanced material developments. Advances in the energy sector and the accelerator sector could hinge upon success in the commercial fusion industry.

## The business case for investment to support stewardship of superconducting magnet technology and conductor development

### *Basic science*

The returns on investment in basic research, while widely debated, is generally understood to be significantly positive [21]. An estimate [22] of the economic impact of the Tevatron found that the economic value of >2,000 PhD degrees alone compensated for the $4 billion investment in the facility, and an overall economic impact was calculated as $50 billion (a factor of 12.5).

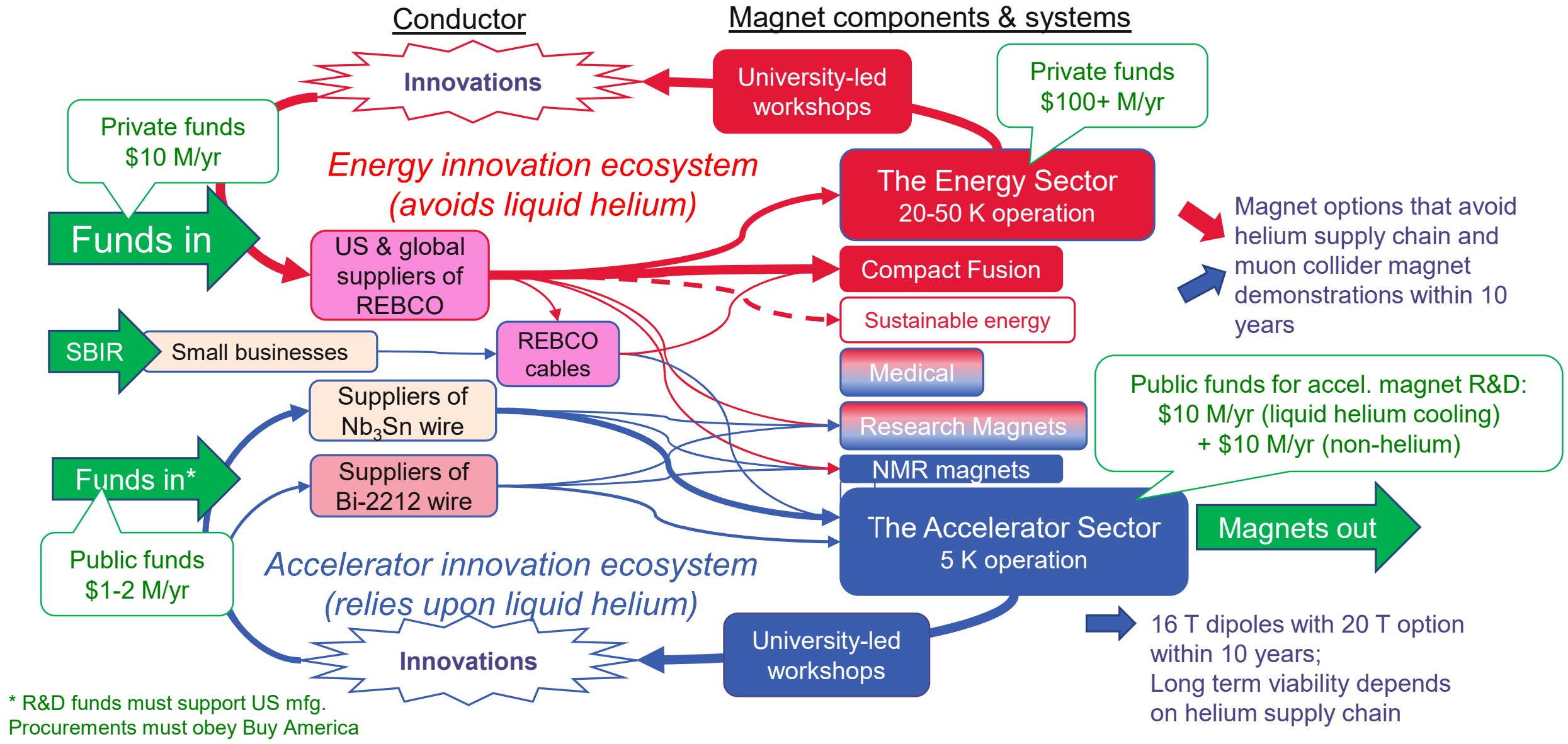

**Figure 6:** Private investment in REBCO outside of the accelerator sector innovation cycle results in significant increase in availability and performance and dramatic cost reduction of REBCO. Importantly, REBCO can be wound into magnets in the form delivered from the factory, whereas $Nb_3Sn$ and Bi-2212 usually require the assembled magnet to undergo a long heat treatment at high temperature. Markets not traditionally coupled to the accelerator sector, in particular markets that favor operation well above the temperature of liquid helium, appear and begin to drive further cost reduction and manufacturing scale. At some point, the traditional conductors used by the accelerator sector could become displaced in the market, potentially creating a supply chain crisis. An early response by the accelerator sector to consider an alternative design paradigm could avoid the supply chain crisis and achieve physics goals in the 2040 timeframe. An opportunity to avoid large stores of helium could also be gained.

Recent history has seen major science facilities such as the Tevatron, the proposed Superconducting SuperCollider (SSC), the Relativistic Heavy Ion Collider (RHIC), the LHC, the High-Luminosity LHC upgrade, and (in fusion) the International Thermonuclear Experimental Reactor (ITER) generate significant conductor production runs that have large disruptive effects on the superconductor industry. The Office of High Energy Physics FY2021 budget indicated expenditures of ~$200 million on science facility construction projects, with comparable forecasted expenses continuing into the next decade. Operations of the facilities above supports many users, e.g. 917 at RHIC and > 6,000 at LHC with >2,400 being PhD students or post-doctoral scholars. While the FCC mentioned earlier might be an extreme possibility, physicists envision a wide range of experiments, each with different requirements on magnets and conductors. During 2023, physicists will participate in the next iteration of the decadal survey to advise the Particle Physics Project Prioritization Panel (P5) about the next large accelerators and the transformational discoveries that could be achieved.

The National Science Foundation likewise invests in large science facilities, including the National High Magnetic Field Laboratory (NHMFL). Magnets for NHMFL users are solenoids or split solenoids, in contrast to the saddle and racetrack-like magnets used in the accelerator sector. The NHMFL is arguably farther ahead in terms of developing HTS magnet technology because solenoids do not require complex bends and mechanical structures used for accelerator magnets. During 2023, the National Academies will convene a decadal panel to review the status of science at high magnetic fields, which will advise about the science and industrial cases for magnet technology. NHMFL hosted 1,615 users in 2021, of which 743 were PhD students or post-doctoral scholars, with funding support of $61.1 million. The projected return per invested dollar is $6.57 to the state of Florida alone. Approximately 80 PhD degrees are generated per year, indicating a contribution to economic growth of ~$160 million.

### *Commercial markets for magnets using accelerator-sector conductors*

**Magnetic resonance imaging:** The MRI global market had a 2021 estimated value of $4.8 billion with a compound annual growth rate (CAGR) of 7.1% [23], with the CAGR of the $250M China sector being as high as 10%. MRI systems are predominantly 1.5 T magnets with some 3 T magnets, with research systems reaching to 11 T on body and 17 T on the coils. The National Academies identifies 20 T MRI of the human head as a grand challenge [6].

**Nuclear magnetic resonance (NMR) spectroscopy:** NMR spectroscopy is an essential tool to probe structure and function of molecules in biology and chemistry. HF and UHF NMR systems provide structural information about proteins and diseases, as well as the structural and functional response to pharmaceuticals. NMR is also essential for understanding bonding, structure, and dynamics in advanced materials for applications such as sensing and quantum information. Since resolution improves at a rate faster than linear with magnetic field increase [24], UHF systems enable access to information not possible to obtain by other means, which creates opportunities for high return on investment. The global market is estimated at $925 million with a CAGR of 3.5% [25]. Higher growth is forecast for the $235 million market portion related to HF and UHF systems. At present, Bruker Biospin is the only supplier of 800 MHz to 1.2 GHz systems with 18.8 T to 28.3 T fields that sell for $2M to $15M or higher.

**Accelerated particle therapy:** Proton and carbon ion irradiation has gained acceptance for its "magic bullet" properties, by which cancerous tissue can be destroyed with negligible impact on surrounding tissue. Therapy systems are more prevalent in Europe than in the United States. A therapy system requires a beamline of protons to be guided by magnets, and cyclotrons to generate and accelerate the particles. Superconducting cyclotrons using high field magnets can be compact and lightweight enough to permit retrofit of existing urban hospital space, which is a competing business model to greenfield sites with larger low-field cyclotron and guiding magnets. The global market is expected to scale from $677M in 2022 to >$1B by 2027 according to Ion Beam Applications (IBA) in Belgium, representing a CAGR of 8.2% [26]. Mevion Inc., in Boston, produces leading-edge technology with 50 employees and a market revenue of $80M. IBA identifies present operations of 250 therapy rooms and a need for 2500 additional rooms in Europe alone. Cost of operations and cost of therapy presently obstruct this potential market.

**High-field measurements systems for academic research:** Researchers in academia and labs operate thousands of research systems to probe behavior of small samples in magnetic fields and other environments. Revenue reports from Quantum Design (San Diego) [27] and Oxford Instruments (U.K.) [28] indicate that the size of the market for scientific instruments with superconducting magnets is about $300M annually. A 25 T measurement system, which is a current goal of product development, might sell for $2.5-5.0 million.

### *Potentially disruptive commercial markets*

**Fusion:** In the absence of fossil fuels and nuclear fission for large scale electricity generation, fusion and hydroelectric become the choices for multi-GW centralized electricity production. Tokamak fusion reactors rely heavily upon advanced superconductors, and there are presently two pathways. The first pathway has been developed using high-field magnets operating in helium at 4–5 K. ITER, presently under construction and scheduled for burning plasma experiments in the 2030s, required 600 tons of $Nb_3Sn$ conductor of a different architecture and lower capability than the conductors used by the accelerator sector [29]. An envisioned DEMO reactor for pilot-scale power production could require 2,400 tons of conductor after 2040 [5], with a conductor cost alone of over $4 billion using present costs.

The second pathway envisions using REBCO conductors operating at 20 K and reaching to much higher fields, 20 T. This pathway takes advantage of scaling factors [30],[31] to reduce the size and potentially the cost of a power plant. The $4.8 billion global private investment in REBCO-based fusion during 2021 envisions electricity production for about $0.025 per kW-hr [31], which is a factor of ~2 lower than investment in renewables (e.g. wind at the present $0.05/kW-hr [32]) for the equivalent energy production. Conductor manufacturing has scaled from < 1 ton/yr to > 3 tons/yr over the past 5 years to meet fusion demand, with scaling foreseen toward > 50 tons/yr. Information publicly available from the U.S. Energy Information Administration shows that electricity generation in 2019 was about 4.2 trillion kW-hr with a market of approximately $1.4 trillion.

**Wind turbines:** Wind turbines represent the leading edge of a broader market in rotating machinery and transportation, which includes generators, motors, ship propulsion, aircraft propulsion, and linear motors in rail transportation. Challenges accompany the use of superconducting magnets for time-changing fields, and the low frequency, < 10 RPM, of turbine rotation presents the lowest technology requirement. General Electric are presently in pilot development of 15 MW class turbines that represent a replacement of permanent magnet systems [33]. Approximately 200 tons of conductor would be required per GW of generating capacity. If REBCO or other HTS materials can be used to generate 6 T field coils at ~30 K to attain 2.5 T or higher across the air gap, then potential to transform the market arises because the power output of a turbine roughly doubles compared to what can be attained with iron or permanent magnets, but the mass of nacelle on top of the tower stays roughly constant, thereby undercutting the present generator mass-on-tower scaling, 20+ tons/MW, and ameliorating a fundamental technology roadblock [34]. The global wind turbine market stands at $85 billion in 2022 and will nearly double by 2030 with a CAGR of 6.3%. Total wind power capacity exceeded 700 GW in 2022 and is expected to exceed 2 TW by 2030. Most growth will be in offshore turbines, where economics favor spending capital to achieve the highest power per tower that is deployable.

## Summary

The return on investment to the US economy for a $1 billion science facility operating over a 20-year lifetime can be expected to be $10-$20 billion, which alone covers an investment of $400 million in the R&D programs under scenarios 3 and 4. Additional hundreds of millions to billions of dollars in return is possible by enhancing existing industries in the U.S. and making it more competitive to on-shore elements of large multi-national companies.

While research supporting this report did not investigate market scenarios in sufficient detail to more deeply examine the growth of markets that require superconducting magnets and conductors important to the accelerator sector, large growth can nonetheless be expected from the markets analyzed above. Below are plausible returns for some US companies that would be enhanced by scenarios 3 and 4:

- Bruker OST, Carteret NJ, might expect a return of $92 million from the NMR market by 2032 based on present estimated delivery of 3.5 tons per year of accelerator-grade $Nb_3Sn$.
- Bruker OST might expect a return of $88 million from the medical therapy market by 2032 based on present estimated delivery of 2 tons per year of accelerator-grade $Nb_3Sn$

- Bruker OST might see the establishment of its Bi-2212 product in NMR magnet prototypes by 2025 and $20 million of production by 2030.
- Mevion, Boston MA, might expect annual revenue to increase to over $150 million by 2030.
- While not directly impacted by the accelerator sector, investments in manufacturing stability will benefit the MRI sector:
  - Bruker OST could see present estimated annual $20 million revenue for MRI production double to $40 million by 2032
  - Luvata, Waterbury CT, could see present estimated $25 million annual revenue double to $50 million
  - Siemens Healthineers USA, Cary NC headquarters, could see its present $3.5 billion medical imaging segment double by 2032.
  - Philips Healthcare USA, Latham NY, could also see doubling of $3.8 billion by 2032.
  - GE Healthcare USA, several locations in NY, SC, MA, likewise could see significant gains in its estimated $6 billion MRI segment double by 2032.
- Quantum Design, San Diego CA, could see the emergence of a new product line starting in 2025 and potential generation of $40 million new revenue by 2030.
- SuperPower Inc., Glenville NY, could greatly benefit from manufacturing improvements to scale from ~200 km annual sales worth $8 million to ~5,000 km annual sales worth $150 million by 2030.

## Implementing the new business plan: Expand the HEP magnet and conductor development programs

The stakeholder discussions identified that existing programs, the Magnet Development Program (MDP) and the Conductor Procurement and R&D program (CPRD) described earlier in this document, are positioned well to carry out an improved business plan. Strengths and challenges for expanding the programs are outlined in Table 1. The elements below constitute a consensus for a baseline program with several options to address stakeholder concerns.

### *Baseline need: $30–40 million per year for magnet and conductor development plus sustained support of university programs*

Scenario 3 with additional investment to mitigate risks identified under Scenario 4 provides the farthest-reaching business plan with potential tipping-point benefits. Expanding the present MDP and CPRD from a LARP-era baseline to cover magnet development roadmaps for $Nb_3Sn$, Bi-2212, and REBCO yields a program of at least $25 million. Escalation for 5% annual inflation over a decade yields $40 million. Operating under this baseline with parallel activities for the multiple conductor options could be expected to proceed at a development pace like that achieved by LARP, with ~2 major milestones achieved per year.

The LARP-era conductor investments also need to be scaled to cover the entire conductor portfolio. For the HL-LHC $Nb_3Sn$ conductor, the 10-year LARP-era R&D investment in RRP conductor alone was just under $4 million,, which would require at least $7 million in 2022 dollars, and this does not include earlier development. So, under the baseline need, CPRD would place *at a minimum* annual orders for 2–4 M$ of conductor, including development conductor at 0.5 M$/yr *per conductor type*, and production conductor at 1–2 M$/yr. This would allow the conductor inventory to keep up with increased demand from magnet development. Manufacturers should be able to rely upon the annual order.

University groups are a crucial part of the accelerator sector. They do not receive funds directly from development programs like MDP and LARP, and instead they compete for ~5 M$ under the GARD

| Table 1: Strengths and challenges for an updated business plan within MDP and CPRD | |
|---|---|
| **Strengths** | **Challenges** |
| • CPRD has an existing charter, mission statement, strategic roadmap [35], [36], management structure, advisory committee, purchasing agent, and methodology for receiving proposals and awarding funds.<br>• CPRD has name recognition and connection to past beneficial outcomes.<br>• Proposals for development activities come from the manufacturers and are vetted by a committee of subject matter experts. While feedback from the committee is often returned, the original innovation comes from industry, and the intellectual property (IP) stays with the industry.<br>• CPRD already maintains a small conductor inventory, and shares its database, numbering, and transaction methodology with other conductor inventories. CPRD would be well positioned to manage a limited conductor stockpile or repository.<br>• CPRD already maintains conductor specifications and quality plans. CPRD would be well equipped to develop specifications and quality plans for new conductors and coordinate conductor testing with university groups and national laboratories.<br>• CPRD has begun to place contracts from SBIR/STTR companies to expand innovation.<br>• CPRD distributes conductor to national laboratories and university groups for research in the accelerator sector. Distributions generally do not require payment to CPRD to replenish conductor. | • CPRD generally does not centrally coordinate or direct research. A business plan to accelerate development of alternative $Nb_3Sn$ conductors and Bi-2212 by coordinating or directing research from CPRD would require significant changes to its charter, oversight, vision, and mission.<br>• CPRD does not fund university groups.<br>• CPRD does not operate facilities that could be used to make or test conductors. CPRD does facilitate the provision of conductors from its inventory for tests and experimental projects.<br>• CPRD does not generally re-sell or distribute conductor from its inventory to industry. CPRD has considered policies such as allowing immediate withdrawal of conductor from the CPRD inventory for distribution to a company when there is a procurement contract already in the company's hand to replace conductor at the company's expense at a future date. Lead time for conductors can be 12 to 18 months. |

university solicitations. Steady support of university groups has provided characterizations, testing, and fundamental underpinning of conductors that innovations rely upon. University-led forums such as the Low Temperature Superconductor Workshop (LTSW) bring together the important stakeholders on an annual basis to identify technology gaps and motivate work to close gaps. This provides effective program review of CPRD and augments the annual external review of MDP. The university portfolio must be sustained and expanded if possible. Moreover, university grants were cited by stakeholders as the most potent source of talent. Since university PhD thesis work often extends across grant renewal cycles, and much longer than SBIR/STTR and other collaborative work agreements, mechanisms to extend grant awards to ~5 years could be beneficial. Mechanisms to build in traineeships could be also accommodated by university centers and institutes.

### *Baseline roadmap:*

The revised MDP and CPRD roadmaps [36] lay out magnet technology and conductor development goals over 2-, 5-, and 10-year time. Under baseline funding these goals become readily achievable, including the outcomes below:

By 2 years:

- Multiple designs of advanced $Nb_3Sn$ advance through research and development stages (see Appendix 1 for discussion about conductor technology advancement).
- Questions associated with Bi-2212 raw powder production are answered, leading to powder specifications and reliable quality assurance.
- CPRD builds a wire inventory that supports magnet development without supply chain delay. CPRD's wire characterization data are publicly accessible.
- CPRD works out accounting methods to provide conductor from its inventory to companies as a strategic reserve when supply chains are especially difficult. For example, CPRD could release conductor from inventory when provided a contract for conductor manufacture that replenishes the inventory from the industrial partner.
- Multiple manufacturers eligible for procurement under rules for use of public funds begin to supply $Nb_3Sn$ conductor.
- REBCO round cables are sourced from multiple vendors in lengths long enough for test magnets.
- REBCO conductors are purchased from multiple manufacturers and characterized to understand the potential for transformation away from round-wire conductors.
- Commercial magnet builders set up agreements to use large infrastructure, testing facilities, and expertise at national laboratories and universities.
- Funding agencies work out how to award grants to universities for ~5 years to better support PhDs.
- Potential risks and benefits of helium-free magnet technologies in the accelerator sector are assessed

In 5 years:

- MDP finds the practical limits of $Nb_3Sn$ magnets with present RRP conductor. The requirements for advanced $Nb_3Sn$, which is needed to extend the practical limits, are solidified.
- $Nb_3Sn$ cables and magnets using conductor from multiple manufacturers are made and tested.
- Advanced $Nb_3Sn$ conductors can be purchased in 40 kg full billets. Performance variations may be wide.
- Bi-2212 conductors can be delivered in 20 kg batches with performance variations less than 10%. Procurements have option to use a specification and quality plan.
- Viability of Bi-2212 for accelerator and commercial magnets is established. A hybrid magnet is made with a Bi-2212 insert and a $Nb_3Sn$ or Nb-Ti outsert.
- Viability of REBCO round cables for accelerator magnets is established. Field homogeneity measurements are reported for the first magnets.
- Fusion companies work with the accelerator sector routinely for magnet, cable, and conductor testing.

In 10 years:

- MDP demonstrates a 16 T dipole using advanced $Nb_3Sn$.
- MDP demonstrates a 20 T hybrid dipole using HTS.
- Advanced $Nb_3Sn$ and Bi-2212 are available as full production conductors with quality plans and specifications. Orders depend on magnet pull.
- Scale round REBCO cables toward pre-production, depending on magnet pull.

### *Option: Expand roles of MDP and CPRD to address other stakeholder needs*

The stakeholder community identified several additional activities that MDP, CPRD, and the university groups should take to address major challenges:

(1) Facilitate increased access to national laboratory and university infrastructure for magnet construction, conductor and magnet test facilities, and design and modeling capabilities. Existing frameworks for cooperative development and technical service agreements are sufficient to stimulate initial interactions.

(2) Bring commercial magnet designers into accelerator sector projects to cross-fertilize designs, techniques, training and best practices, materials, supply chains, workforce, and other important aspects of building magnets and systems.

(3) Facilitate coordination between accelerator sector needs and commercial needs to arrive at common specifications, quality plans, testing needs, and measurement science.

(4) Develop training programs for workforce, possibly in conjunction with establishing a national traineeship center at locations where pipelines of students and other talent presently exist. Candidate locations include NHMFL and the Applied Superconductivity Center in the Florida A&M University – Florida State University College of Engineering, the Center for Superconducting and Magnetic Materials at Ohio State University, the Texas Center for Superconductivity at the University of Houston, and the Accelerator Technology Center at Texas A&M University.

(5) Explore a national manufacturing center of excellence or a manufacturing innovation institute connected with university and national programs to facilitate process development, innovations at small scale, transfer of ideas from research to larger development, and exploration of new raw materials.

### *Option: Additional support of a sprinting environment for more rapid development of advanced $Nb_3Sn$ and Bi-2212 round wires*

The CPRD roadmap calls for vigorous development of advanced $Nb_3Sn$ and Bi-2212 to achieve production conductors in 5-10 years. Even with increased funds for conductor development, the organizational structure of CPRD is not well suited to lead focused development. Both advanced $Nb_3Sn$ and Bi-2212 have challenges that require university groups, industry innovators including SBIR companies, and large industry manufacturing to come together to address challenges and their integration along the production path. It could be possible to re-organize CPRD in such a way that larger allocations of funds are provided to manufacturers, where manufacturers then follow up with R&D agreements with universities. This model worked successfully under the DOE Office of Electricity Superconductivity program at the $10M to $40M scale. For smaller development effort, support for central coordination could be a better option.

University groups have already supported development by coordinating procurements of raw materials and injecting ideas drawn from basic research into manufacturing. **An additional investment of $1-2M annually in university centers could augment the present activity by allowing the formation of**

**university-led consortia.** This could work in harmony with the present CPRD model, where proposals for innovation come from manufacturers, because the university centers in the accelerator sector have their own conductor fabrication facilities that work in concert with their extensive characterization and testing facilities. Stakeholders thought that the university groups' ability to see the connected challenges and direct work toward solutions could facilitate faster development in the initial stages where both advanced $Nb_3Sn$ and Bi-2212 are now. Sabbaticals for commercial project leaders at universities could allow for exploration on university research equipment followed by direct transfer to industry development equipment. Internships and interactions by academic and laboratory leaders directly on the industrial fabrication lines could reveal unknowns and focus activity on problems at early manufacturing stages. Once developments that mature sufficiently, they could then graduate from the consortia back to the CPRD model.

It could be possible to combine different conductor types under a single consortium, or to form separate consortia based on material type. Because ideas flow from the academic teams to manufacturers, the universities would be responsible for intellectual property agreements. Stakeholders in manufacturing favor non-competing models and royalty-free licenses.

### *Option: Building a limited stockpile of magnet conductors*

Continued demand creates a strong incentive for conductor manufacturing to remain in the United States. *Stockpiling* a limited amount of conductor smooths supply-chain variations and provides an on-demand inventory for accelerator sector magnet R&D and potentially also the commercial magnet ecosystem. Supply chain delays are presently 12 to 18 months due to raw materials and specialized facility scheduling (e.g. extrusion services). A stockpile or national repository will need to make careful prioritization of *production* conductors used in magnet engineering while metering in, when appropriate, *developmental* conductors whose reproducibility or reliability falls short of the requirements for engineering. Stakeholders felt that **RRP conductors should be stockpiled at a level of $1M/yr** which is equivalent to 500 kg or 100 km of Hi-Lumi strand. They also felt that **an initial priming investment of ~$5M would be prudent**. The Hi-Lumi quadrupole magnets required 500 m of 40-strand cable (20 km of strand) to manufacture each 4.2 m long, 2-layer coil, i.e. 8 layer-meters. This same "unit length" would supply cable to one research dipole magnet with typical length of 1 m and 4 layers per coil (2 coils needed per dipole). The stockpile investment above would thus support 5 research magnets per year. The stockpile will need to carry a variety of strand types to accommodate the multiple cable types needed by magnets. Bi-2212 strand can also be stockpiled at a smaller level since to accommodate strand architecture changes slowly over time. A $1M/yr investment would support a 60 kg addition to the inventory, which would supply 12 km of 0.8 mm strand and allow research magnets to scale up in size. A stock of REBCO round cable could also be prudent if initial magnet tests are fruitful.

### *Option: Additional support to encourage formal connections between the accelerator sector and the potential REBCO-driven market by including muon collider research*

Among the large projects being accomplished by the accelerator sector is a muon collider. Design studies for a muon collider are active now, as outlined by several contributions to the particle physics "Snowmass" process [17]. Magnets for a muon collider include approximately 2 km of solenoids ranging in field up to perhaps 50 T. Many of these require HTS technology to be applied in various ways, where the upper field range corresponds to grand challenge goals for high magnetic field science [6] set by the National Academies. **Support for augmentation of programs under MDP and CPRD, or investment in a dedicated program, at the scale of ~$10 million annually could provide the following benefits and risk mitigations:**

(1) The accelerator sector becomes more closely integrated with the magnet technologies presently being developed or considered for REBCO. This includes aspects such as:
    a. Winding strategies, pancake or layer.
    b. Insulation strategies, which vary between traditional turn-by-turn insulation to "no-insulation" approaches.
    c. Strategies for mechanical structures, since REBCO is an anisotropic conductor and imparts atypical forces to magnets.
    d. Strategies for quench detection and protection, where REBCO can operated in temperature ranges where propagation of hot zones is orders of magnitude slower than for typical magnets in the accelerator sector.
    e. Strategies for instrumentation.
    f. Strategies for cooling, including helium-free conduction cooling.
(2) The accelerator sector gains more direct relationships with multiple conductor manufacturers, adding to the manufacturers who presently supply a very special conductor architecture for the manufacture of round cables. This includes aspects such as:
    a. Understanding the different conductor architectures, manufacturing methods, opportunities for scaling, conductor costs, and selling price.
    b. Understanding physical differences between each manufacturer's finished product, including dimensional control, slitting edges, surface finish, and artifacts of handling and any quality control.
    c. Understanding structural differences between different manufacturers and how these differences affect magnet operation.
    d. Understanding the different electromagnetic properties across different manufacturers. Recent experience shows that variations are wide across manufacturers [37] and even for a single manufacturer [38].
(3) The accelerator sector gains opportunities to explore magnet designs that accommodate the anisotropic shape of REBCO to investigate basic questions such as whether any approach can provide the needed field homogeneity.
(4) By engaging with the REBCO community, the accelerator sector becomes part of the transition and will be much more able to monitor its impact on the present round-wire technologies.

## Appendix 1

*Reprinting of a 2022 white paper prepared for the Accelerator Frontier sub-panel reporting to the Particle Physics Project Prioritization Panel (P5), "Challenges and opportunities to assure future manufacturing of magnet conductors for the accelerator sector." arXiv preprint arXiv:2208.12379 (2022).*

(this page left blank intentionally)

# CHALLENGES AND OPPORTUNITIES TO ASSURE FUTURE MANUFACTURING OF MAGNET CONDUCTORS FOR THE ACCELERATOR SECTOR

White Paper for the Accelerator Frontier
Snowmass '21

Lance Cooley[1,2], David Larbalestier[1,2], and Kathleen Amm[3]

1 Applied Superconductivity Center (ASC), National High Magnetic Field Laboratory (NHMFL), Florida State University
2 College of Engineering, Department of Mechanical Engineering, Florida A&M University and Florida State University
3 Superconducting Magnet Division, Brookhaven National Laboratory (BNL)

## Executive Summary

***$Nb_3Sn$ magnet conductors will continue to be the workhorse material for accelerator magnets over the coming decade because they can deliver significantly higher magnetic fields than Nb-Ti at significantly lower cost than higher performance HTS conductors. High current-density $Nb_3Sn$ conductor suitable for present accelerator magnets for the High-Luminosity LHC (Hi-Lumi) upgrade, i.e. ~12 T field, is commercially produced in long lengths in a mature fabrication process. R&D enhancements for "advanced $Nb_3Sn$" could improve performance by 30% or more in the 15-16 T field range envisioned for future dipoles, and conductors could become available in 2–5 years. New developments in cabled REBCO tape HTS conductors and 2212 round-wire strand are creating opportunities for hybrid $Nb_3Sn$-HTS dipole magnets approaching 20 T and solenoid magnets pushing toward 50 T. Disruption of Nb-Ti and $Nb_3Sn$ magnet technology and increasing cross-over points for hybrid and stand-alone HTS options operating above 4 K temperature should be expected as HTS magnet technology continues to develop and mature.***

***No superconductor is presently manufactured at the huge scale required for a high energy or muon collider. MRI Nb-Ti is the only present tonnage conductor, but its architecture is simpler than any envisaged accelerator conductor. Moreover, it is made at very small margin because of continuing large cost pressures from the MRI magnet makers. The superconductor industry of today lacks adequate economic margins from Nb-Ti MRI wire manufacture to support development of innovations such as advanced $Nb_3Sn$ and HTS conductors. Market pull for the Hi-Lumi grade of $Nb_3Sn$ exists is presently insufficient to drive manufacturing scale beyond the 30-ton Hi-Lumi procurement.***

***Accelerator magnets require premium grades of $Nb_3Sn$ conductor that have a limited supplier base. For example, the $Nb_3Sn$ conductor used for the High-Luminosity LHC has only one supplier that can meet the required performance and piece length, whereas the larger procurement activity for ITER was sourced from 8 suppliers. With Hi-Lumi coming to an end, there is significant concern about how to keep premium $Nb_3Sn$ manufacturing active enough to prevent loss of expertise and atrophy of capabilities. It is our view that present public-private partnerships must be enhanced. Annual procurements of significant scale and conductor stockpiling will be necessary to maintain pre-production readiness of existing conductor designs. Re-thinking the innovation network will be needed to facilitate conductor innovation in view of the tension between large-volume, low-margin tonnage Nb-Ti and the smaller-scale, Hi-Lumi $Nb_3Sn$ with limited end-use pull, and developmental $Nb_3Sn$ and HTS conductors which cannot be funded by present industry profit margins.***

***Broader coordination between federal offices and the conductor-magnet industry ecosystem could encourage the emergence of marketplace demand that results in both better and lower cost conductors. Synergies are envisioned between a sustained development program for a muon collider and technology maturation for emerging industries since both entities require high-field solenoids. Increased partnership between accelerator sector magnet developers and magnet technology for industry could provide important bridges across development obstacles by making national laboratory infrastructure available to companies. The reshaping of national agendas to address climate and energy sustainability is instigating new applications in fusion and electric machines with cross-fertilization to accelerators. Transformations in health and medicine also have potential overlap with conductor and magnet technology in the accelerator sector. A portfolio of industries could be nurtured by the accelerator sector via innovation institutes and magnet technology centers of excellence to serve national and economic security needs and advance future accelerator technology needs.***

## Development path of magnet conductors from small samples to large-scale production

### *Basic requirements at small-sample scale*

Superconducting materials that could eventually become important conductors start out from discoveries with small samples, e.g., pellets, thin films, small cast or melted rods, etc. Among the starting points of today's conductors are arc-melted alloys of Nb, sintered powders of $MgB_2$ and $(Ba_xK_{1-x})_2FeAs$, and films of $YBa_2Cu_3O_7$ (YBCO, or more generally REBCO where RE= Y or other rare earth element). Measurements of certain basic properties reveal whether the material has potential to impact magnet technology. Among the first tests of a new material are the determination of basic attributes including:

1. Whether there is a high enough upper critical field or irreversibility field at accelerator operating temperature (usually 4 K). See Figure 1.
2. Whether high macroscopic (across many grains) and microscopic (within one grain or crystal) critical current density $J_c$ is present. See Figure 2.

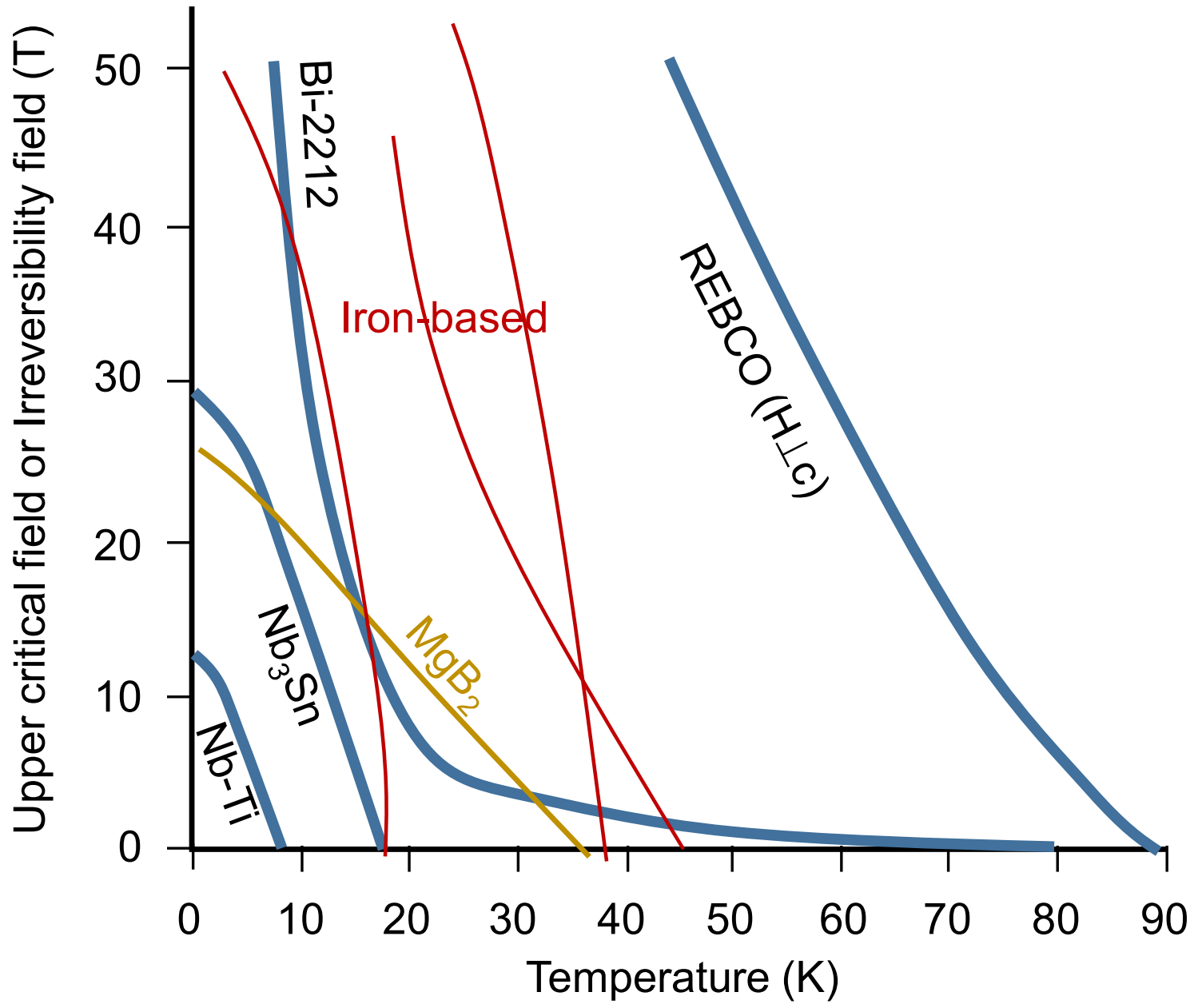

**Figure 1:** A plot of upper critical field or irreversibility field vs. temperature for magnet conductors used by the accelerator sector and potential new conductors.

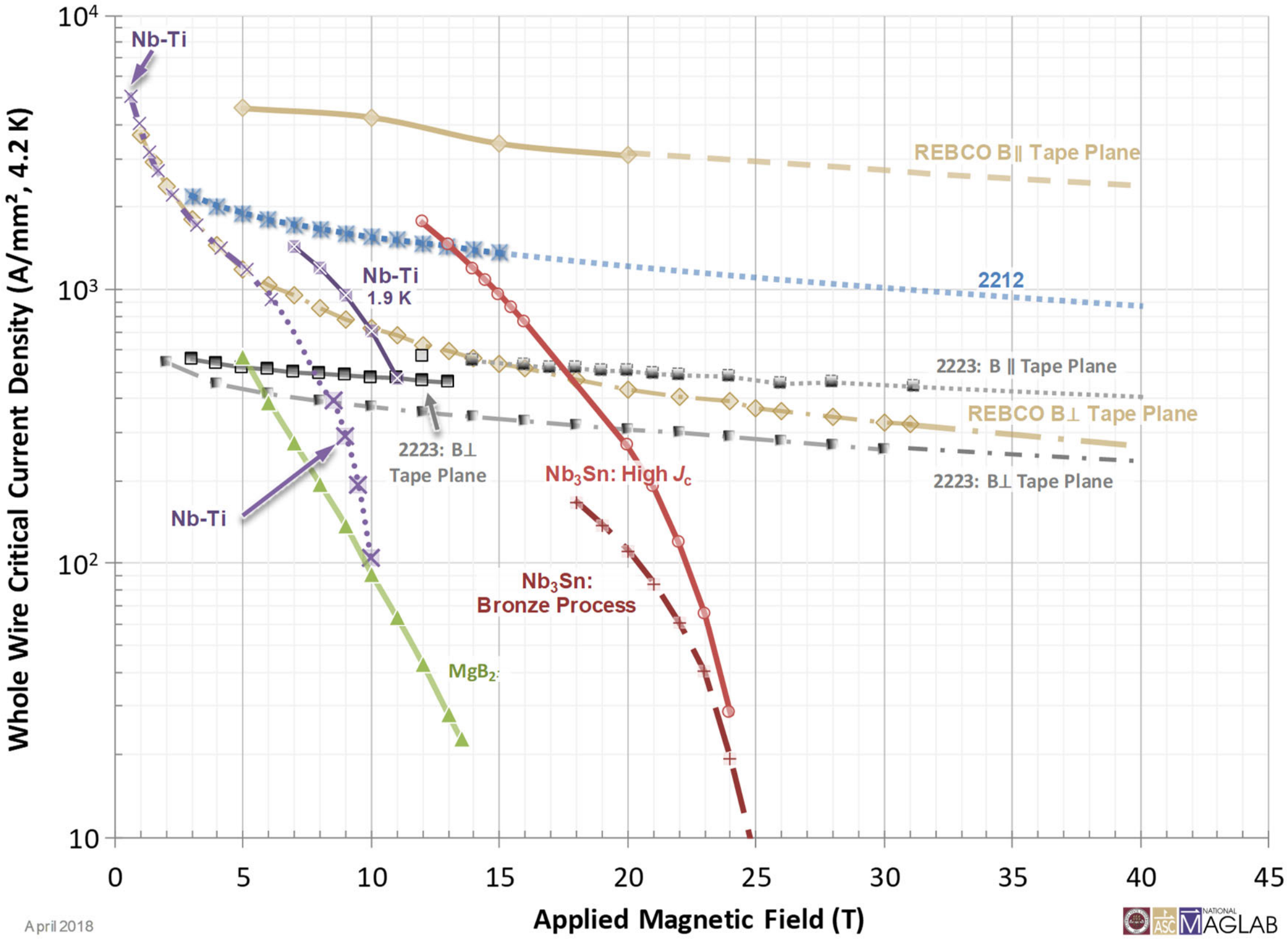

**Figure 2:** Critical current density of industrial wires and tapes used in superconducting magnets. Plot is maintained by Peter Lee, see https://nationalmaglab.org/magnet-development/applied-superconductivity-center/plots.

3. When blockage of current is evident, understanding the cause of the connectivity obstruction must be attained, so that methods of enhancing that connectivity can be developed (for example, by melt texturing of Bi-2212 ($Bi_2Sr_2CaCu_2O_8$) or the biaxial growth of REBCO on textured templates).
4. Whether the synthesis methods are compatible with parallel-bonded high conductivity metals, especially copper.
5. Whether there is significant crystalline and property anisotropy, which may strongly impact vortex pinning and mechanical properties.
6. Whether there are ready means to increase flux pinning by suitable additives and mitigate connectivity limitations by for example grain boundary engineering.

Statement: Support of materials science research that allows the HEP community to capitalize on new materials and enables their application to provide technological breakthroughs is and has been vital to the development of superconducting applications.

## *Considerations for research on nascent wires*

Research to develop a new material in a wire or tape form takes place using billets or stacks that are up to about 2 kg mass and lengths ranging from a few meters to 100 m. These ranges are tied to the capabilities for development in university, SBIR, and industry research settings. To perform the first stages of wire development, requirements *in addition to those for basic material studies above* are:

1. Compatibility with copper wire fabrication methods, i.e. wire or tube drawing to large strain. *Composite* wire manufacturing approaches include combinations of ductile metals (e.g. Cu + Nb-Ti alloy, Cu + Nb + Sn for later reaction to form brittle $Nb_3Sn$) and combinations of packed powders in ductile tubes (e.g. Cu or Cu-alloy tube + $MgB_2$ powder, Ag tube + Bi-2212 powder, possibly Cu tube + powder of Fe-based superconductor).
2. If not, then compatibility with long-length fabrication by other means, e.g. reel-to-reel deposition on (REBCO and its buffer and template layers on Hastelloy or stainless steel) or printing. The 2D conductors emerging from such processes complicates magnet technology.
3. Understanding how processing serves to define properties is the basic materials science paradigm that drives all optimization. The materials science involved in the discovery of a promising superconductor can be very different than the materials science that goes into production of viable magnet conductors capable of being fabricated in long lengths. All processing steps must serve the optimization of structure and thereby optimization of properties—this was the key lesson from the development of Nb-Ti magnet conductors.
4. Development of heat treatments and other reaction processes that are compatible with magnets. Essentially all materials with operational envelopes beyond Nb-Ti are brittle, meaning that they must be made from ductile components that are placed into a furnace to allow the superconductor to form *in-situ* via reaction, e.g. $3Nb + Sn \rightarrow Nb_3Sn$. To avoid cracking of the brittle material, magnets are usually fabricated by "wind and react" methods, which implies that the conditions for the superconductor formation reaction must also be compatible with the materials used in the magnet. This includes compatibility with insulation and with mechanical and structural components. $Nb_3Sn$ and Bi-2212 are generally used in a wind-and-react approach.
5. Understanding of how the materials depend on strain, both reversibly and irreversibly. These properties must be understood at the basic superconducting materials level and at the conductor level where normal metal conductor such as Cu or a strong substrate like Hastelloy may be present. The strain limits must be determined at the whole-conductor level, and these may enable alternative "react and wind" magnet strategies, where the conductor is reacted to form the brittle superconductor on a spool and subsequent magnet winding avoids strain above the threshold for cracking (REBCO is always supplied in this final superconducting state).
6. If the superconductor is not isotropic and is formed with texture on a tape substrate (e.g. REBCO), then understanding of the variation of properties with orientation is essential.

### *Considerations for industry R&D and scale-up*

Transfer of nascent wire ideas to industry comes with expectations of a product and viability of a marketplace. Trials scale to the size of partial and full billets, for instance 10, 20, and 45 kg for $Nb_3Sn$ corresponding approximately to 1–10 km single wire piece lengths. Basic attributes of a product need to emerge from industrial R&D, including:

1. Compatibility with existing manufacturing infrastructure to avoid unnecessary new investment in infrastructure that would generate a wider "valley of death" before adoption.
2. Understanding requirements for specialty manufacturing. Some manufacturing paths require techniques that do not overlap the art of copper wire manufacturing. Examples include the special $NbSn_2$ powder manufacture developed for powder-in-tube (PIT) $Nb_3Sn$, vapor deposition methods, specialty shape forming, and additive methods. Specialty manufacturing requirements can be a significant cost detriment, as for example for PIT $Nb_3Sn$.
3. Flexibility of conductor architecture and opportunity for innovation. Innovation cycles of "better, faster, cheaper" greatly help both users and manufacturers.

4. Scaling to long length with high yield, low breakage, and low rejection rates. Product margin must cover capital, operating, and materials expenses associated with unsellable pieces not suitable for customer use.
5. Consistency of superconducting and mechanical properties along the length of single pieces, from piece to piece, and from production batch to batch.
6. Consistency of key performance parameters such as critical current along the length of single pieces, from piece to piece, and from batch to batch. For $Nb_3Sn$, conductor optimization has at least two performance parameters: critical current and copper conductivity.
7. Development of measurement and testing capabilities needed to unambiguously validate key properties.

### *Considerations for industry production readiness*

The full production run for a future accelerator facility could require procurement of hundreds or thousands of tons of magnet conductor over a period of several years. Production readiness requires:

1. Determination of production specifications and related quality assurance.
2. Demonstrated ability to deliver to specifications with low reject rate.
3. Successful delivery of orders of significant size sustained over long duration. Annual procurements of 100–1000 kg were sustained over a decade during LARP [1], establishing manufacturing capability and generating much-needed production and delivery statistics. Similar procurements preceded the major production runs for the LHC, ITER, SSC, and RHIC.
4. Sufficient information to arrive at a procurement cost estimate.
5. Availability of test facilities and inter-laboratory benchmarks or standards to verify the above.

| |
|---|
| Statement: Driven by magnet development needs at national laboratories, collaborations between fundamental materials researchers in universities and labs with the conductor manufacturers has produced a 40-year long virtuous cycle of continuous conductor development for the accelerator sector. Sustained support for this “secret sauce” is vital to optimization of future conductors too. |
| Statement: Sustained large procurements of conductor are vital to keep manufacturing capability warm, encourage retention of know-how and workforce, validate expectations for a production run, and encourage growth in the related non-research marketplace. |
| Statement: Discussion about conductor cost should not begin until production readiness has been achieved and supply chains are established. |
| Statement: Support of standards and test facilities is crucial for industries to quantify performance and assess technology and manufacturing advancement. |

## Magnet conductors in the Accelerator Sector portfolio for the next decade

Table I summarizes in approximate terms the conductors available for procurement in 1 km lengths that meet the requirements of accelerator magnets. For example, $Nb_3Sn$ conductor designs for the High-Luminosity LHC upgrade (“Hi-Lumi LHC”) are referenced but those for fusion are omitted, because fusion designs are optimized with parameters unsuitable for high-field accelerator magnets. The conductor development life cycle described above occurs over ~1 decade of time, so the position of $Nb_3Sn$ as the workhorse conductor and HTS as contributing conductors should only be expected to change slowly. We comment about potential disruptive factors, which could change the table substantially, at the end of this

| Table I: Magnet conductors available for procurement in length > 1 km | | | | |
|---|---|---|---|---|
| SC material | Billet or batch mass | Annual production | Relative cost | Comments |
| Nb-Ti | 200-400 kg | Hundreds of tons | 1 | Driven by MRI industry |
| $Nb_3Sn$ RRP | 45 kg | 5–10 tons | 5 | Driven by general purpose and NMR magnets and by Hi-Lumi LHC |
| $Nb_3Sn$ PIT | 45 kg | < 1 ton | 8 | Cheaper RRP is also generally more capable |
| Bi-2212 | 20 kg | < 1 ton | 20–50 | See note (1). |
| REBCO | 10 kg | < 1 ton; few tons for fusion | 20–50 | See note (2). |
| Bi-2223 | 20 kg | < 1 ton | 20–30 | Current leads (3). |
| $MgB_2$ | 20 kg | < 1 ton | 2 [5] | Current transfer cables feeding magnets. |

Notes:

(1) Costs for Bi-2212 are artificially high because full-time R&D teams are supported by only a few conductor orders per year.

(2) Privately funded fusion projects, which target e.g. 20 T at 20 K, envision magnets that do not require large stores of helium. This could make REBCO the choice for the accelerator sector if avoidance of helium becomes a priority.

(3) Although Bi-2223 (i.e. conductor based on $Bi_2Sr_2Ca_2Cu_3O_{14}$) has current density that is at present significantly lower than that of either Bi-2212 and REBCO, Bi-2223 is unique for bridging temperature zones from 110 K downward, such as for magnet current leads.

document. Several $Nb_3Sn$ conductors under development, e.g. PIT and Hf-alloyed variants, have the potential to reach 1 km length and would be expected to have manufacturing profiles like the production conductors shown below. A comprehensive review of development activity led by CERN was published recently [2]. $MgB_2$ and Bi-2223 conductors are included because of their potential role in supporting accelerator magnets, e.g. like has been done for the CERN Superconducting Link [3], and in niche magnets where heat load is important [4].

Statement: End-use pull is the most significant factor that drives down cost and motivates production scale-up. End-use pull from marketplace products is weak for $Nb_3Sn$ compared to Nb-Ti. Commercial HTS magnet technology is just beginning and synergistic development with the accelerator sector is now starting, especially through SBIR and STTR vehicles.

### *Restacked-rod-process RRP $Nb_3Sn$ conductor: the present workhorse conductor*

The conductor produced for the Hi-Lumi LHC, shown in Figure 3, uses an architecture called "restacked rod process" [7],[8] wherein monofilament rods of Nb or Nb alloy are sheathed in copper, extruded, drawn, and stacked in an annulus around a central Cu core, wrapped with a diffusion barrier, and extruded a second time. Part of the core is then replaced with tin, and this assembly forms what is called a sub-element. After further drawing, the sub-elements are stacked (i.e. the original rods are re-stacked) and further drawing proceeds to the final size. A reaction sequence produces mixed Cu-Sn phases at low to moderate temperatures (below 400 °C) before a final high-temperature segment facilitates a diffusion reaction between Cu-Sn and Nb (or Nb-alloy) to form $Nb_3Sn$ (or $Nb_3Sn$ alloyed with Ti, Ta, etc.) [9].

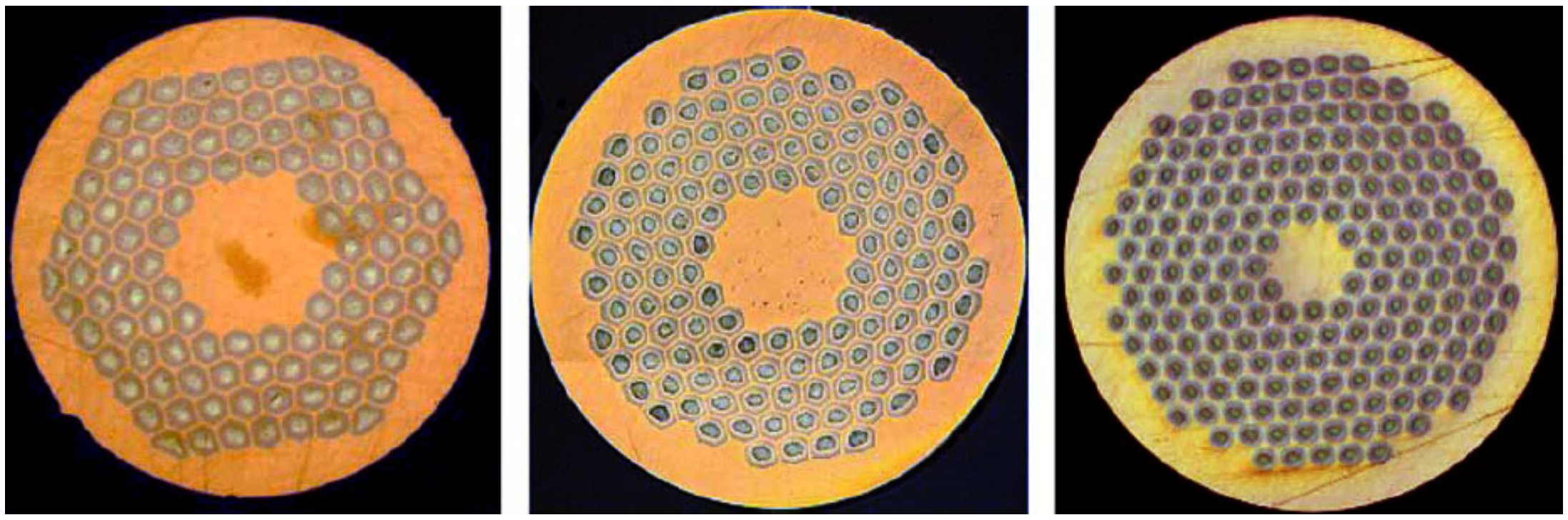

**Figure 3**: Cross-section of different 0.85 mm diameter Hi-Lumi LHC quadrupole conductors: At left is a 108/127 RRP strand from Bruker-OST, center is a 132/169 RRP strand from Bruker-OST, and at right is a 192/217 PIT strand from Bruker-EAS. Images were supplied by the manufacturers to the upgrade project as part of quality control. See Ferracin *et al.* [6].

Conductor naming typically refers to the number of sub-elements that occupy a theoretical number of sub-element sites while keeping hexagonal symmetry. The theoretical number obeys the formula $3n(n–1) + 1$ for $n$ layers around a central rod (the first layer), i.e. 1, 7, 19, 37, 61, 91, 127, 169, 217, 271, … for $n$ = 1 to 10. The number $n$ also identifies the number of hexagonal units along the flat outer edge of the stack. The Hi-Lumi quadrupole strand used 108 of 127 possible sites, and so was identified as "108/127", with the central 19 sites being copper. Research strands leading up to the Hi-Lumi LHC project explored configurations restacks from 61 up to 217 sub-elements, including copper [10],[11]. Procurements in 2021-2022 specify 150/169 and 162/169 designs and 1.0 to 1.1 mm final diameter.

A key limitation on reducing hysteretic losses and minimizing field errors due to large magnetization currents is presented by the fact that the final reaction stage grows the individual Nb filaments into a solid $Nb_3Sn$ annulus that makes the effective filament diameter $d_{eff}$ that of the whole connected annulus rather than the starting Nb filament unless countermeasures are taken to divide the annulus, e.g. by substituting pure Ta rods [12]. Here, the product $J_c d_{\text{eff}}$ is associated with energy contained in the strand magnetization, which can be uncontrollably released as a flux jump and potentially initiate a magnet quench. The effective filament diameter is approximately given by

$$d_{\text{eff}} = d_w[N(1 + R)]^{-1/2}$$

where $d_w$ is the wire diameter, $N$ is the number of sub-elements and $R$ is the ratio of stabilizer copper area to total sub-element area. Flux jumps are serious challenges for $d_{\text{eff}} > 60$ µm. For the 0.85 mm diameter Hi-Lumi quadrupole strand, $d_{\text{eff}}$ is nominally 55 µm. Re-stacking a larger number of sub-elements to make $d_{eff}$ smaller, increasing $N$, however, works in opposition to manufacturing yield, uniformity, performance, and cost. These trade-offs presently dictate present RRP design optimizations. Cables for future dipoles using present RRP designs will need larger diameter strands to carry more current, with present procurements seeking 1.0 to 1.2 mm diameter. Keeping $d_{\text{eff}}$ at ~60 µm requires re-stacks with $N$~200. Higher $J_c$ $Nb_3Sn$ conductors would allow smaller wires, adding flexibility into the RRP strand design and also the fabrication of cables made from them.

The RRP design has proven to be quite flexible in over a decade of development. The tin insertion step and the stacking pattern in the annulus provides the means to vary the ratio of Nb to Sn, thereby tuning the $Nb_3Sn$ reaction dynamics. Higher Sn content, e.g. a Nb:Sn ratio of 3.4:1, creates higher tin activity and drives the average $Nb_3Sn$ composition across the layer closer to stoichiometry for a typical final reaction stage, e.g. 665 °C for 48 hours. This results in higher irreversibility field and $J_c$. However, aggressive reactions can drive Sn through the diffusion barrier and into the copper. Tin is a strong electron scatterer in copper, and even as little as 0.1% Sn can reduce the conductivity, as measured by the residual resistance ratio RRR, by a factor of 10 [13]. The resulting contamination adversely affects quench protection. A reduction in the availability of Sn, by using Nb:Sn ratios of 3.5:1 to 3.6:1, results in better ability to aggressively drive the diffusion reaction with less risk of reducing RRR [11]. This gives better multi-parameter optimization [1]. Figure 4 shows the outcome of this change for the Hi-Lumi LHC production run, which used a ratio of 3.6:1.

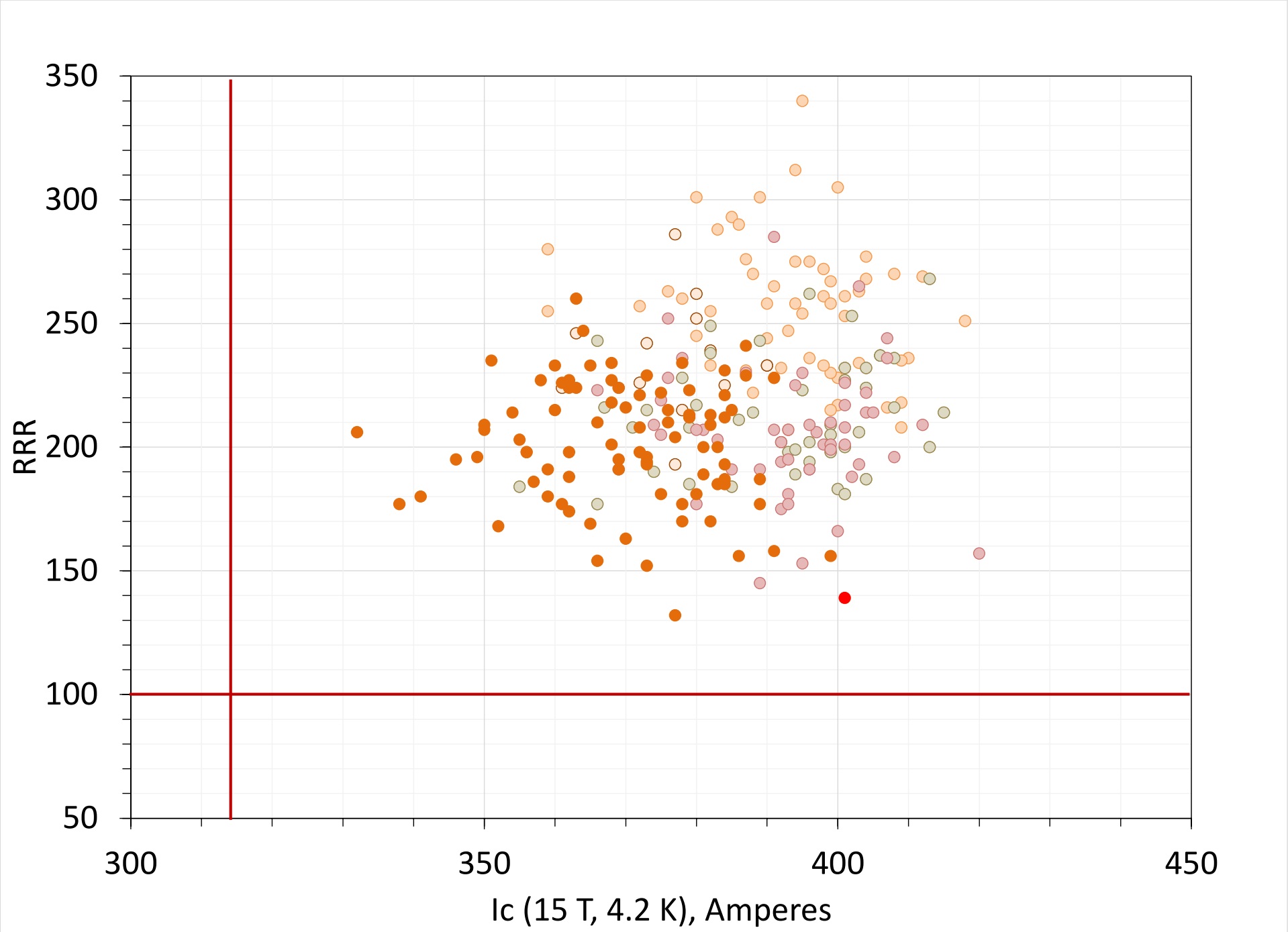

**Figure 4:** Cumulative data for RRP strand delivered for the US contribution to the Hi-Lumi LHC upgrade is represented in this plot of the RRR vs wire critical current at 15 T, 4.2 K. Each dot represents a quality control sample tested by the manufacturer, and the different shading identifies different procurement stages over the entire run. Importantly, these data are for strand rolled to reduce the diameter by 15%, which was found to simulate very well the degradation at sharp bends in Rutherford cables. Vertical and horizontal lines define the specifications for rolled strand.

Adjustments of the architecture provide resilience against damage due to cabling. The process to make Rutherford cables requires rolling through a Turk's head where sharp bends at the cable edges lock in the strand geometry. Cabling research, which is not discussed at length in this document, typically probes the limit of compaction where strands are locked into position, but the performance is not seriously degraded. The RRP architecture facilitates some degree of adjustment to accommodate deformation of sub-elements during cabling. However, excessive local shear deformation of sub-elements can break diffusion barriers and allow Sn to have unobstructed access to the copper, usually resulting in significant loss of RRR. For Hi-Lumi quadrupole strands, rolling of round test strands to 85% of the initial diameter produced properties that tracked well with the degraded properties observed from strands extracted from cables, where in particular the RRR of the rolled strand was less than that measured around the sharp bends in cables [14],[15]. Rolled strand experiments thus provided a conservative predictor of cable performance. Also important is that the critical current of RRP strands is quite robust against rolling or cabling, because for reasons not fully understood the Nb-Sn diffusion reaction converts nearly as much of the sub-element area to $Nb_3Sn$ in distorted sub-elements as for sub-elements free from distortion.

The continued improvement of RRP conductors will be vital to future colliders. Two sections below address activities that crosscut all $Nb_3Sn$ conductor designs: (a) replacement of Pure Nb filaments by Nb-alloys; and (b) incorporating high heat-capacity materials into the strand architecture. In addition to these opportunities, improvement of RRP conductors also needs research attention given to:

- Understanding better how the sequence of Cu-Sn and Cu-Sn-Nb phases that evolve during the reaction heat treatment might be optimized for better $Nb_3Sn$ performance at >15 T. Prior work [16],[17] identified opportunities that that have not yet been fully exploited.
- Reducing $d_{eff}$ without concomitant loss of $J_c$ performance, loss of yield, or increase of cost.
- Scaling production to 100 kg billets or larger.
- Modifying the conductor architecture and heat treatment to optimize properties above 15 T.

### *Powder-in-tube PIT conductor and rod-in-tube RIT conductors*

Tube-based $Nb_3Sn$ approaches replace the annulus of stacked rods in RRP conductors with a solid tube of Nb or Nb-alloy. In principle, this removes one stage from the fabrication process, i.e. RRP: monofilament Cu/Nb, sub-element annulus, finished conductor; PIT: packed mono-tubes, finished conductor. The tin source, often combined with copper, can be a powder or a rod inserted in the center of the tube, whereby a radial diffusion reaction is used to form the $Nb_3Sn$ layer. The assembled tube and core material is restacked in a similar fashion as described for RRP conductors, where design flexibility is provided by tube shape (hexagonal or round), separation, and composition of components in the tube core. Reports [18] investigated restacks with as many as 744 tube sub-elements ($n$ = 18 with 169 sub-elements being Cu) at tube diameter approaching 15 μm for 0.7 mm wire diameter.

Tubes are sourced as a raw material, which means that conductor manufacturing must extend from the art of composite wire drawing to also encompass the art of seamless tube manufacture. Piercing plates followed by deep drawing or spinning, dynamic flow forming, hydroforming, electroforming, and seamless mandrel extrusion are established methods for tube manufacture. Tubular conductors can in principle be made both with powder inside the tube [19],[20] as in the PIT conductors made by Bruker-EAS using $NbSn_2$ powder [19], or with solid rods, as for the conductors rod-in-tube conductors made by HyperTech Research [20]. More than a decade of development both for NMR magnet and for Hi-Lumi have shown it to be a flexible architecture but also one which is ultimately inferior to RRP architectures. An important mechanical concern is that all powder conductors require powder sliding during wire fabrication and this cannot be accomplished with full density cores. Typical residual porosities are about 1/3 which leads to greater irreversible strain sensitivity than in full density RRP wires [21],[22],[23]. A second factor is that the reaction path from the $NbSn_2$ source to $Nb_3Sn$ goes through "Nausite" $Nb_{0.25}Cu_{0.75}Sn_2$ and $Nb_6Sn_5$ which forms as a very large grain phase that rejects Cu into its grain boundaries when finally converting to very large grain $Nb_3Sn$ with poor vortex pinning, impaired connectivity and low $J_c$ properties [24],[25],[26],[27]. Thus the highest $J_c$ values of PIT have always been inferior to those of RRP conductor when expressed in the important conductor metric of $J_c$ equal to the critical current divided by the area of all phases including the diffusion barrier needed to make the $Nb_3Sn$.

In addition to these important areas, and in addition to the research topics discussed for RRP conductors, further improvement of tube-type $Nb_3Sn$ conductors could result from:

- Improved manufacturing experience with tubes
- Improved manufacturing of fine powders of Nb-Sn and Cu-Sn intermetallic compounds
- Development of full density routes using ductile Cu and Sn components
- Continued use in magnets to reveal conductor vulnerabilities

### *Other $Nb_3Sn$ designs*

A significant advantage of the RRP and PIT conductor designs above is the placement of tin, generally as a Cu-Sn phase, next to the niobium in sub-elements. This creates high tin activity, which leads to rapid reaction and smaller composition gradients in the $Nb_3Sn$ that yield higher $J_c$ in the layer and higher irreversibility field averaged over the whole A15 phase layer. Some conductors produced for ITER incorporated a large central source of tin or used bronze as the Sn source. The reduced tin activity of α-bronze works against achieving the higher layer $J_c$ properties of RRP and PIT, while the low solubility of Sn in α-Cu Sn bronze means that the fractional cross-section of A15 that can be formed with bronze conductors is about half that in RRP and PIT. Thus, high Sn conductors develop high $J_c$ both because of better layer properties and a higher volume fraction of the A15 in the overall reaction mixture.

Interesting exceptions have been summarized by Hopkins [28],[29] working in conjunction with CERN and manufacturers in Japan and Russia. Distributed-tin designs could advance after extended R&D to approach performance targets for high-field magnets. Continued support of research into alternative designs and in manufacturing capabilities worldwide will be required to alleviate supply chain challenges if production for future facilities is on the scale of that completed for ITER (which sourced from 8 manufacturers) and well beyond the scale of the Hi-Lumi LHC upgrade (which sourced from 1 manufacturer). Consolidation of manufacturing since the ITER procurement has reduced the overall manufacturing capacity and capabilities, however. At present no new architecture yet challenges RRP for overall conductor $J_c$, scale and supplied piece length. In our opinion, one of the biggest challenges of all high Sn routes is the complexity of the initial Sn-Cu mixing and then the various kinds of reaction layer that form while diffusing Sn form its source into the Nb.

### Advanced Nb alloys for $Nb_3Sn$ conductors

The metals in Groups IVa, Va (Nb is in group Va), and VIa form body-centered cubic alloys with significant mutual solubility. This condition permits adding elements that have strong affinity for oxygen (e.g. Zr, Hf) as well as elements with high atomic number (e.g. Ta, Hf, W). Development of $Nb_3Sn$ tape over 30 years ago by General Electric [30], [31] utilized Nb1%Zr alloy coupled with anodization to create *in-situ* oxidation, by which $ZrO_2$ particles constrained $Nb_3Sn$ grain growth during a subsequent high-temperature reaction. Virtually all $Nb_3Sn$ conductors use grain boundaries as vortex pinning centers to drive up $J_c$. In addition to the much greater electromagnetic stability of bronze multifilament conductors, the ability to form the A15 phase at 650-700 C ensured that grain sizes of 100-200 nm were possible, not the micron-sizes of tapes formed above 930 C.  Tapes disappeared from the market shortly after low temperature reacted, multifilament bronze conductors appeared [32] due to their excellent electromagnetic stability and much higher $J_c$. When accelerator builders took up the challenge of making $Nb_3Sn$ suitable for accelerators, it was realized that bronze conductors could not generate sufficient critical current density and this stimulated much of the internal tin architectures that form the precursors for today's RRP & PIT conductors. Dietderich *et al.* [32] broadened the discussion by noting that oxide particles could not only constrain A15 grain growth but also form strong vortex pinning centers within $Nb_3Sn$ grains. Under the stimulus of the challenges of an Energy Frontier Hadron Collider, a number of researchers [34], [35], [36], [37], [38] have subsequently explored the potential benefits of using internal oxidation to improve flux pinning and increase $J_c$ at high field in $Nb_3Sn$ conductors. Conductor designs to achieve this goal will be discussed shortly.

The benefits of enhanced vortex pinning would be lost without concomitant retention or increase of the upper critical field. Here, addition of elements with high atomic number provides spin-orbit scattering [39], [40], [41] and suppression of any paramagnetic limiting, thereby enhancing the upper critical field. Tarantini *et al.* [42], [43] show the opportunities for increasing the upper critical field of $Nb_3Sn$ conductors when alloyed with both Ta and Hf.

Because a central goal of all $Nb_3Sn$ conductor development is to grow the A15 phase with grain size significantly smaller than 100 nm, these recent efforts focused attention on how the nucleation of the A15 phase occurs at the Nb-Sn reaction front. An important new observation in [37] is that addition of 1at.%Hf to Nb4at%.Ta raises the recrystallization temperature of the generally used pure Nb rods in standard RRP designs above that of the normal A15 reaction temperature. It appears that A15 nucleation occurs by diffusion of Sn into the alloy GBs, which allows A15 grains as small as 50 nm even in the absence of any oxide particles which confer similar benefits [34,35,36]. This new combination of grain refinement techniques is under active study at the present time.

Development of conductors by both RRP and tube routes is presently underway using advanced alloys in both university and national lab and industry. These developments demonstrate well the presence of a collaborative ecosystem linking initial discovery in universities with “nascent” wires made in university or small business to quasi-industry R&D. Greater conductor development resources could greatly accelerate the ability to ensure that real magnet conductors become available in the next 3-5 years. Special Nb alloy melting has been addressed by multiple raw material suppliers, and fabrication activity also spans multiple industry participants.

### *Additions of high heat capacity materials into conductors*

Research from the Kurchatov Institute and Bochvar Institute in Moscow more than a decade ago demonstrated the ability of rare earth additives to add stability to magnet windings [44], [45] that could accelerate the rate of magnet training [46]. More recent considerations [47], [48] identify several challenges of $Nb_3Sn$ magnets that could be alleviated by the addition of rare-earth oxides. As discussed earlier, the sub-element diameter of present RRP and PIT conductor types fall into a regime where magnetization instabilities can be a significant factor in initiating training in present $Nb_3Sn$ dipoles. This challenge will become more difficult if $J_c$ is further increased using advanced conductor designs if $d_{eff}$ remains the same. Addition of a material with high heat capacity could offer an important stability counterbalance to the potential increase of sub-bundle magnetic energy by helping absorb and slow the energy released during flux jumps. Magnet stability and training can be addressed at the strand, cable, and coil level by further additions of materials with high heat capacity, e.g. via additives to epoxy or additions to the cable core. These are activities outside of the scope of this white paper.

Development of $Nb_3Sn$ conductors with high heat-capacity additives is underway in parallel with investigation of advanced alloys mentioned above. Challenges exist in identifying and then optimizing powders for both properties and compatibility with manufacturing.

Statement: $Nb_3Sn$ conductors with advanced designs and improved performance at high field are presently in R&D. Pushing them into industrial pilot plant production so that magnet lengths of wire can be made within the next 3-5 years and will allow proper test of the magnet potential.

### *Bi-2212 round wire conductor*

Conductors made from high-temperature superconductor (HTS) are capable of much higher fields than any Nb-based conductor, as Figure 1 shows. Bi-2212 is quite unique as a cuprate HTS material in being able to develop high $J_c$ in a multifilament round wire form without macroscopic texture. Bi-2212 offers the potential to displace $Nb_3Sn$ above about 15 T field, see Figure 2, in round multifilament wire form. A key advantage of Bi-2212 over REBCO and Bi-2223 is that the processing that develops high $J_c$ passes through a state in which starting Bi-2212 powder melts to an almost complete liquid phase. Careful resolidification under conditions of sparse nucleation allows rapid growth of a few grains in the confined Ag filament tubes [49]. All cuprate superconductors require texture to permit current to pass from grain to grain; in the case of Bi-2212 made by powder-in-tube methods the melting step allows disconnected Bi-2212 phase powders that deform inside their Ag tubes by particle rolling to form connected dense liquid regions on melting. The problem of the melting though is that the residual 1/3 volume is principally air that suffers an ~4-fold increase in pressure that can swell or even burst the very soft Ag sheath. The oxygen can pass through the Ag but this is only 20% of the gas pressure since nitrogen remains trapped. During the slow solidification process that ensures sparse nucleation, large grains form randomly with low density. Fortunately, the 3 crystallographic axes have significantly different growth rates. Within filament tubes of ~15 μm diameter, the most rapid *a*-axis growth gobbles up off-axis grains and generates a strong [100] texture along the filament axis. The *b*-axis is orthogonal to *a* thus generating a biaxial texture of ~15°. A key issue of this aligned microstructure is that obvious weak link connections are suppressed, unlike Bi-2223, where only a uniaxial texture exists [51] [49], [50], [51].

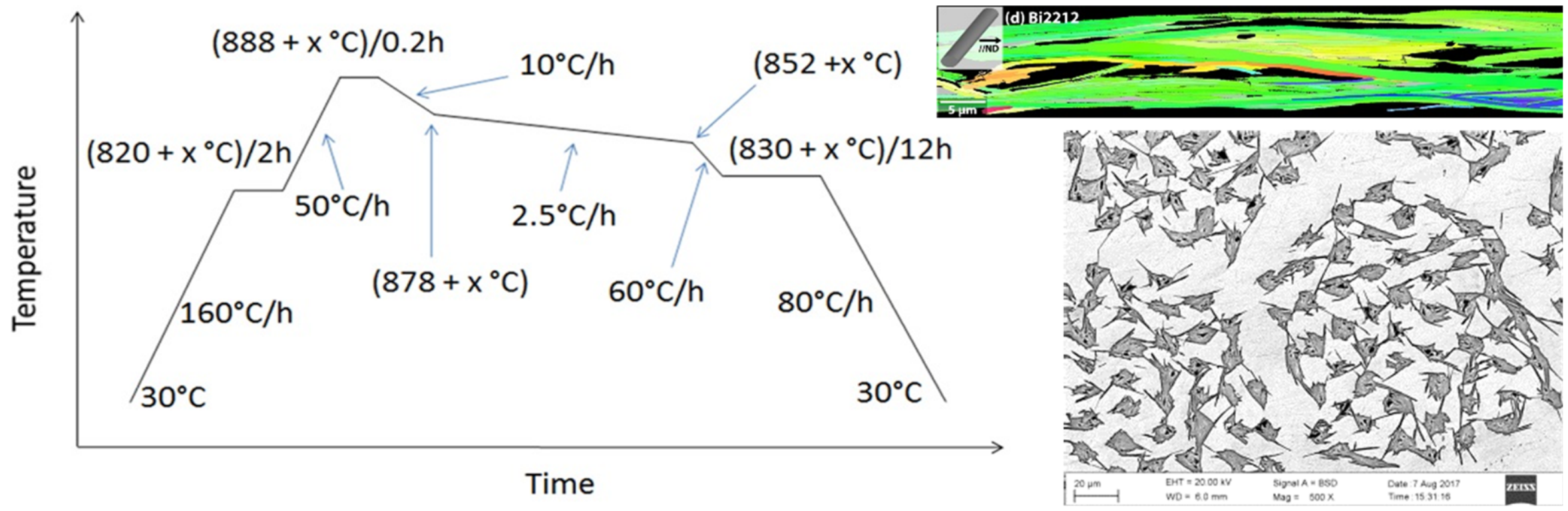

**Figure 5:** The melt-solidification process sequence and the resulting microstructure of Bi-2212 strand. A key aspect, partly related to the crystal structure of Bi-2212, is that crystal growth during slow re-solidification is much faster along the crystallographic *a* direction, which lies within the copper-oxide planes. When the filaments are well-defined and separated, as in the bottom right photo, this leads to the formation of macroscopic texture and alignment of the copper-oxide planes, as represented by the predominance of green color in the orientation-dependent image of a longitudinal cross section at top right. Texture is essential for avoiding obstruction to carrying current when anisotropic superconductors such as Bi-2212 are used.

The second important step in obtaining high $J_c$ is remedy the effects of entrained air in the powder and its tendency both to burst the wire during heat treatment (HT) and divide the filament into full density regions and voids or bubbles across which current flows only by bridging the voids by a few rapidly growing large Bi-2212 grains [51], [52]. The remedy is to perform the HT under sufficient pressure to prevent any effect of the internal gas pressure [53], [54] that would prevent full densification of the Bi-2212 filaments. Operation over the range 20-100 bar total pressure (1 bar is always oxygen with balance argon) allows >95% density of the Bi-2212. Because 2%$O_2$ in argon is standard welding gas, we have standardized to this formulation also for 50 bar overpressure HT.

Figure 5 summarizes the HT, the texture, and the final filament structure of an optimally processed Bi-2212 wire after such a 50 bar over-pressure HT. Such an OPHT for coils necessitates a wind-and-react magnet approach in specially designed apparatus at Florida State University shown in Figure 6.

A round wire HTS architecture is important for the accelerator sector in several ways. First, wire-drawing techniques presently applied to copper-clad superconductors can be applied to Bi-2212 in large degree. This could permit scale-up and quality control like that essential to large-scale production of Nb-Ti and $Nb_3Sn$ conductors for accelerator magnets. Second, cabling techniques are generally unchanged between Nb-based conductors and Bi-2212. Third, although the need for twisting is not fully clarified, twisted conductors have been produced. This could provide an important advantage for ac loss in rotating machines, with possibly greater impact than the need to reduce losses upon ramping of accelerator magnets. Fourth, existing insulation winding and braiding schemes can be adapted to accommodate the temperature and chemical interactions for Bi-2212.

Many of the basic requirements for understanding Bi-2212 conductors and obstacles to their use in magnets were worked out by the Very High Field Superconducting Magnet Collaboration between 2009 and 2013, which combined HEP national laboratory teams with university groups. Further advances were achieved by advancing the production of Bi-2212 powder between 2013 and 2018 with a standard composition, the so-called "521" formulation [55]. Multiple, small-scale, manufacturers have produced good, fine powders that facilitate conductor fabrication with low breakage. Continued support from both DOE-HEP and the National Science Foundation (NSF) via the National High Magnetic Field Laboratory (NHMFL) has led to demonstrations of good reproducibility for coil reactions based on conductor witness samples.

New over-pressure reaction systems should allow dipole model coils up to 1 m length and 250 mm diameter to be completed by end of 2024, if provided adequate resources. This infrastructure should also facilitate exploration of laboratory solenoids and NMR systems at well above 30 T field. Infrastructure for longer magnets has received initial informal consideration.

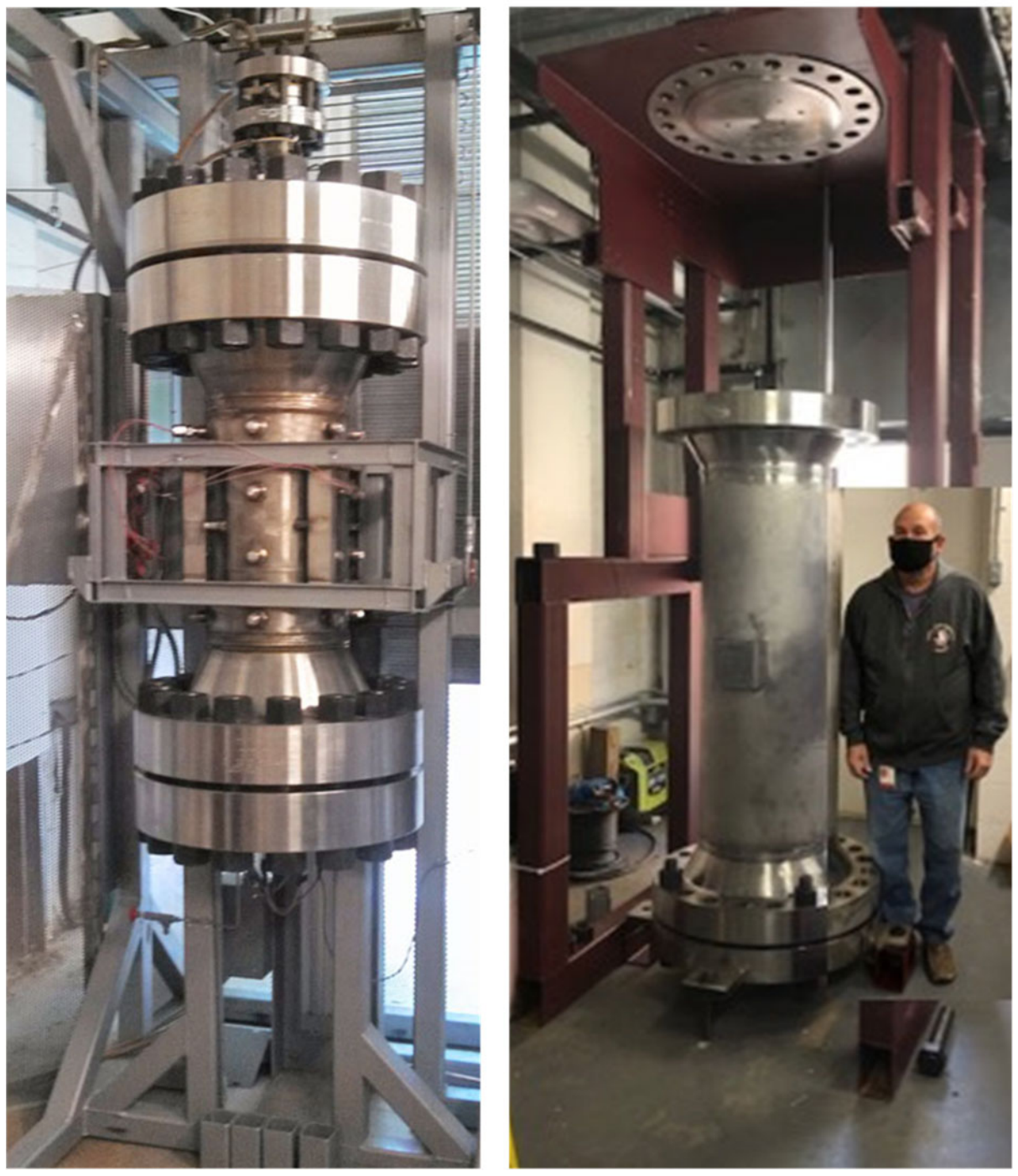

**Figure 6:** Reaction furnaces for Bi-2212 test coils operating at 50 bar pressure through the reaction sequences like that shown in Figure 5. Dimensions reflect the internal diameter and length of the uniform temperature zone. The larger furnace at right is being commissioned in 2022.

### *REBCO conductor*

REBCO became a viable high-field magnet conductor when high-strength Hastelloy substrates became dominant [56], [57] and when advanced pinning center design showed how to develop extremely high layer current density in manufactured tapes [58], [59]. REBCO rapidly became the basis for high field magnets, especially the 32 T user magnet at the NHMFL [60], [61] and the Bruker Biospin 28.2 T, 1.2 GHz NMR system. Numerous other magnet demonstrations above 25 T have now been made as Table II summarizes. Such work, as for Bi-2212, has important bearing on the solenoids for a potential muon collider. Several projects are underway aiming toward 40 T magnets [62].

Further advances to reduce the Hastelloy thickness from a standard 50 µm to 30 µm [63] provide a significant enhancement of the winding current density, especially in the so-called "no insulation" condition [64]. This enhancement led to achievement of winding current densities of over 1400 A mm$^{-2}$ and reaching beyond 45 T field [65]. Stacked-tape cables [66], [67], [68] now make up a vibrant research field for fusion technology and potentially other applications for user and industry magnets. The scope of these manifold developments is too broad to summarize properly in this document.

Roebel cables were first developed for REBCO conductors about 15 years ago [69], [69], and they have been implemented in short model dipole coils [71]. Extensive characterization of Roebel cables [72] uncovered susceptibility to degradation under transverse stress, which could be partly managed by epoxy impregnation. A difficulty with the Roebel design is that wide REBCO tapes are required as starting material, and over half of the starting material is cut out and discarded by the patterning process, significantly driving up the cable cost. These challenges remain at the present time.

Thin Hastelloy is an enabler of compact round "wire" REBCO cables that provide an interesting alternative to Roebel cables. Conductor on round core (CORC®) [73] and symmetric tape round (STAR) [74] are two variants envisioned for future accelerator magnets. Each consists of a REBCO production tape conductor wound in a helical fashion around a conductive core. Thin substrates reduce the strain applied to the REBCO layer [75], which permits winding on smaller forms which enables smaller wire diameter and higher current density. Wires with typical diameter of ~ 3 mm are facilitated by the 30 µm Hastelloy product. Flexibility to bend either CORC® or STAR conductors is related to the tape width and winding pitch, where conductors that are narrower (now pushing below 2 mm width) and wound with a shorter helix pitch have the best flexibility. Unfortunately, these trends drive up cost by requiring more conductor per unit length of cored wire. At the present time losses due to damaged material at the slit edges is not negligible. Indeed development of laser slitting is proceeding fast at multiple companies, making gross damages implicated in magnet damage much less likely [63].

A key question for the future of REBCO coated conductor (CC) has perennially been: *When will the price come down to be attractive for electric utility users*? For more than a decade, the technical success of coated conductors in transmission cables, fault current limiters, motors and generators was thought to be a demonstration of replacement of many copper and iron machines with superconductors. The central dilemma was that coated conductors showed no possibility of getting to acceptable production levels and supposed low cost without commitments that were predictably unrewarding. All of this has potentially changed within the last two years as the case for tokamak fusion power plants operating with magnets with peak fields of 20 T at 20 K has gained huge private support. A prototype toroidal field coil used 500 km (about 1.5 ton) of 4 mm tape, well over 2 orders of magnitude more than the coated conductor put into *all* prior HTS magnets. The SPARC reactor being constructed will require 9,000 km (about 30 tons) more. Potentially therefore, there is now a very important feedback loop in place: Fusion magnet demands keep lowering the price of CC, allowing its replacement use for Cu and Fe for electro-technology. Such cost reductions have great promise for future colliders provided that suitable conductors for accelerator dipoles and quadrupoles can be developed.

Real project use has many benefits one of which is that routine testing generates important information about the stability of production. Such a program for the 11 km of tape purchased for the NHMFL 32 T user solenoid showed that many aspects of one company's production had significant property fluctuations of many types, in $J_c$, in dimensions, in Cu thickness and in damage generated by mechanical slitting. While some of these properties are tracked by the vendor, typically critical current properties are tracked only at 77 K and self-field. A central challenge is that the applications in the accelerator sector require verification of properties at 4 K, high field, and a range of angles, regimes in which it is much more difficult to carry out measurements due to the extremely high critical current and the associated risk for sample damage. Application of the so-called "lift factor" to estimate 4 K performance from 77 K breaks down when contributions from point pinning centers, e.g. oxygen vacancies, add to and compete with the correlated pinning due to artificial pinning centers, e.g. barium zirconate nanorods [76], [77], [78]. Significant variability in the 32 T conductor properties at 4 K have has been noted, even when the 77 K data shows reasonable uniformity [79]. Variations in properties have been connected to variations in the nanostructures, believed to be induced by fluctuations in growth conditions of the REBCO that are apparently beyond present process control both between different production runs and sometimes along the length of a single conductor. Some processing variations can lead to vulnerabilities in the magnet [80], [81]. In a sign that the highly optimized pinning centers possible in laboratory systems may not be replicable in 500—1000 m lengths, one important company has recently described a much simpler chemistry based only on Y, Ba, Cu and O where the dominant pinning centers are $Y_2O_3$ nanodisks generated by controlling the excess Y present during growth [82].

The discussion above points to a need for tight feedback between characterizations at 4 K and processing by the different manufacturers. While several discussions (some referenced above) have looked at the products coming off manufacturing lines, explicit optimization for 4 K and high field is only just beginning. A significant characterization effort is going into the manufacturing of cables, especially for fusion, and the research and development of very high field magnets. Several facilities around the world, notably at NHMFL, Robinson Research Institute, Tohoku University, and University of Geneva, are providing detailed characterizations of the strand for these cables in a field and angular range of interest for the accelerator sector. The vulnerability of cables to stress is receiving new attention in accelerator magnets [71],[83] and some fusion cable tests [84], [85], [86].

### *Iron-based superconductors*

Iron based superconductors (FBS) present the opportunity of a material with very interesting primary properties and potential to become an important conductor. It is now in the delicate transition from the research lab to industry. Routes to make it as simple wires using the powder in tube route have been demonstrated [87], [88] but at present the best current densities are not compelling for applications, being about 100-300 A/mm$^2$ at 10 T 4.2 K.

FBS encompass several compound classes [89], [90], [91]. They are medium temperature superconductors with the interesting compounds having critical temperature $T_c$ lying in the 35-55 K range and extrapolated upper critical fields >50 T at 0 K (see Figure 1). In the rather isotropic so-called 122 compounds, of which $(K,Ba)Fe_2As_2$ may be the most interesting example, critical temperature $T_c$ is ~35 K and $H_{c2}(0)$ as high as 90 T with virtually no $H_{c2}$ anisotropy. The lack of anisotropy and the very high $H_{c2}$ are unequalled by any other superconducting compounds, setting up an immediately strong practical and scientific case for their use [92], [93]. Single crystals can have $J_c$ values exceeding $10^5$ A/mm$^2$ at 4.2 K and a few tesla. Sadly, early experiments [94], [95], [96] with [001] thin film bicrystals showed evidence for weak link behavior beyond about 5° misorientation, albeit with about a factor of 10 less angular $J_c$ sensitivity than in YBCO bicrystals. It is now accepted that there is an intrinsic weak link behavior in FBS.

However, early wires did demonstrate on the scale of magneto-optical imaging resolution, (~5 mm), that current flow was uniform [87] and that transport and magnetization (using the whole sample dimensions) measurements of $J_c$ agreed, all of which argues in favor of there being a true long-range supercurrent. The reasonable conclusion is that there are percolative supercurrent paths through the generally 100-500 nm grains that are much more effective than in comparable polycrystalline REBCO bulk or wire samples.

At this stage there is a strong Chinese program that has publicly announced a belief that FBS, especially $(K,Ba)Fe_2As_2$ can become a viable conductor and the group of Ma at CAS-Beijing has demonstrated 100 m long monofilament wires [97]. Their wires, actually tapes, do exhibit some mechanical texturing in rolling that appears to enhance $J_c$ in the favorable direction, but not yet to a level that is compelling for applications. In summary FBS materials are at the interesting stage where fine-grain polycrystalline compacts can be made as bulks or wires by PIT fabrication routes with current densities that approach the lower level of interest for applications (~100 A/mm$^2$ in 10 T at 4 K). The problem of not-yet-high-enough $J_c$ appears to be one of degraded percolative connectivity. Whether the connectivity is best enhanced by texture, by greater attention to purity (especially of grain boundaries) or managing of internal strain within the conductor (the $T_c$ of all FBS is very sensitive to the Fe-As bond angle) is not yet clear and is being actively studied by multiple groups [97], [98], [99], [100]. In short, the scientific case for researching how to enhance the connectivity of polycrystalline FBS remains strong because the route to wires is quite open if the appropriate materials processing route to higher connectivity can be understood. Moreover, their raw materials are inexpensive and routes to make wire conductors might be compatible with copper composite wire manufacturing.

## The role of public-private partnerships in conductor development

### *Conductor Development Support under DOE-HEP*

Superconducting wire development has been supported by DOE-HEP ever since the Tevatron, initially within the Fermilab, BNL and LBNL programs [101], [102] and then with the University of Wisconsin program [103]. Fermilab showed how to make Nb-Ti at scale for the Tevatron and exposed the fact that wire properties were variable and poorly understood. This attracted the attention of the Wisconsin group who saw the opportunity for improved wires flowing from a proper understanding of the process. The principal vehicle for stimulating the virtuous cycle of better understanding leading to higher $J_c$ leading to better wires was an annual workshop (The Nb-Ti Workshop first held in 1983, still running annually under the new title of the Low Temperature/High Field Superconductor Workshop). This workshop, now generally called LTSW, has always been small, typically 75 persons, to allow an unscripted discussion and interaction vehicle for the magnet users, the superconducting industry and the researchers trying to understand the pathways to better conductors. The Nb-Ti developments generated by this public-private partnership led not only to the SSC, RHIC and LHC conductors, but also to the huge tonnage industry of Nb-Ti for MRI magnets.

The success of this program then led to a new program whose goal was to develop new and much higher $J_c$ $Nb_3Sn$ for accelerator magnets [104], [105], [106]. Under the Conductor Development Program (CDP), large manufacturers supplying Nb-Ti wire for the MRI industry, who were not eligible for incentives like those given to small businesses, were aided to innovate new wire products. Awards were based on proposals for innovative work and industry cost-share. This public-private partnership model led to key innovations of the RRP conductor in the years leading up to the Hi-Lumi LHC upgrade production run [1]. Combined investment of just under $4 million by LARP and CDP between 2005 and 2014 advanced the RRP configuration through 61, 127, and 169 stack variants as well as established production readiness [107]. While the scale of internal matching funds are not available to the public, a good estimate of the total *annual* development investment in high-$J_c$ $Nb_3Sn$ conductor is $1 million. This was followed by pre-production procurements of about 1 ton leading up to the Hi-Lumi LHC upgrade.

Unfortunately, the landscape for public-private partnerships has changed. Margins derived from medical imaging systems have decreased significantly since about 2000, which has practically eliminated the funds available for industrial cost share. The successor to CDP under DOE-HEP, the Conductor Procurement and R&D Program or CPRD, has broadened its scope to support $Nb_3Sn$, Bi-2212, and most recently REBCO, but has not been able to provide resources like those described above leading up to the Hi-Lumi LHC upgrade. As a result, advance of $Nb_3Sn$ with new alloys and additives, as well as advance of Bi-2212 and REBCO as magnet conductors for the accelerator sector, has not proceeded as quickly as the program would like. Significant additional support does come under small business programs grants and research investigator awards, and these have been essential for supporting conductor development through the nascent wire stage.

### *Nurturing the industry ecosystem*

Real challenges lie in moving conductor development from nascent wires into industry R&D and pre-production. Present industry "pull" for high-performance $Nb_3Sn$ wire is ~1 ton annually, which is orders of magnitude below production levels that would be needed for a major science facility. Importantly, the medical imaging market has not taken up $Nb_3Sn$ conductors in a major way, and other markets for therapy and high-field NMR magnets do not generate as large a demand for advanced conductor as the MRI sector. The conclusion of the $Nb_3Sn$ production run for the Hi-Lumi LHC upgrade could mark an important point in time, where investment decisions for $Nb_3Sn$, especially the premium RRP grades sought by accelerators, and Bi-2212 face a long, uncertain period until a possible future accelerator facility is launched, with a very modest marketplace in the interim. (We discuss REBCO in view of private funding for fusion below.) These considerations could force a shutdown of manufacturing in the US and consolidation of resources to other locations. Atrophy of expertise and capability would then follow, which would make the re-start period much longer and more expensive.

Continuity of the CPRD program is presently integrated within the goals of MDP. Actions to sustain and augment MDP and CPRD will be essential to continuing conductor development under the present mechanisms. This model is unlikely, however, to successfully compensate for the huge cost pressures felt by the wire industry from the MRI magnet manufacturers for whom Nb-Ti has become a commodity. It is important to note that the conductors desired by the accelerator sector generally require the highest performance: for example, $Nb_3Sn$ for accelerator use has the highest $J_c$ of all routes. Although today's RRP wires do find application for many types of laboratory magnets, including the highest performance NMR magnet wires [108], manufacturing of strand for accelerators is always aiming for the highest properties that renders their goals somewhat decoupled from manufacturing to meet the needs of other customers. The recent end of the production run for the Hi-Lumi LHC upgrade places urgency on finding a better model for assuring conductor supply for a future science facility. This loss of demand, taken together with considerations such as manufacturing leases, anticipated conductor orders, and federal project profiles, could compel manufacturers to take actions that significantly impair US conductor manufacturing by 2024.

During 2022, supply-chain exercises in the DOE were initiated by an Executive Order [109], which propagated to various workshops and reports [110]. Funds were provided by the DOE Office of Accelerator R&D and Production (ARDAP) to explore the nature of public-private partnerships that could alleviate magnet conductor supply-chain risks and better assure availability of production for the accelerator sector. A workshop held 13-14 March 2022 at Tufts University debated many aspects of a potential business plan going forward. Among the elements thought to be essential for the plan are:

- Sustained magnet development programs like those driven by the accelerator sector, e.g. LARP and programs proposed for consideration by HEP [111]. Magnet development programs provide a virtuous cycle of defining ever more challenging targets and goals, pushing industry to develop ever better conductors in large quantities, making and testing lots of magnets (sometimes to the point of failure), and exposing and solving issues that underpin the advancing technology. Programs with coupling to solenoids and other magnet configurations used by industry applications would especially be beneficial to the accelerator sector – industry ecosystem. R&D toward muon colliders are especially exciting for this reason.
- Sustained annual acquisitions of conductor over and above R&D wires to keep manufacturing “warm” at a pre-production level. Acquisitions of raw material, e.g. advanced Nb-Ta-Hf alloy, needed to allow emerging ideas to enter industrial R&D production could be very valuable and are presently being nurtured at a low level by CDRP. Pre-production levels for $Nb_3Sn$ would be 1 ton per year with configurations meeting standard specifications, e.g. the Hi-Lumi LHC upgrade specification. The configurations should anticipate future use by the accelerator sector, where for instance 1.0 mm diameter wires might be favored over 0.85 mm diameter.
- Managing a limited conductor stockpile as a national resource for both accelerator magnet R&D and growth of the commercial magnet industry. The stockpile would be limited to avoid inventories of obsolete conductors where R&D is rapidly moving. Development of new product lines and valley-of-death development challenges have been traditionally assisted by public funds, especially in technology areas of strategic interest. Assistance could involve withdrawals from the stockpile as prudent. Over the long term, development of market pull should emerge, creating a virtuous cycle like the development of Nb-Ti conductors.
- Nurturing workforce pipelines by supporting student internships in industry settings, industry traineeships at academic institutions, training partnerships, and post-graduate incentives. Besides universities, trade schools and junior colleges are seen as adding journeyman skills to the conductor and magnet manufacturing workforce.
- Coordinating an innovation institute to centralize a pool of manufacturing knowledge and establish a creative common around the key questions for conductor R&D and scale-up. Such an institute would build on the capabilities developed over 40 years by the culture of the Nb-Ti workshop and LTSW/HFSW. Entrepreneurial ideas that have strong merit often do not produce fruit for reasons related to limited resources or technical difficulties. An institute would coordinate access to capabilities and equipment, facilitate exploration of ideas through multiple attempts, encourage debugging of failures, and catalog methods with appropriate propriety. Legal teams involved in partnerships could be encouraged at early stages. Marketing and communication teams could also be engaged during early discussions.
- Incubate component development especially where components, such as superconducting solenoids, provide a strategic or critical advantage for larger systems. Products and components could be listed on the federal register.
- Incentivize industry ecosystem collaboration between national laboratories, universities, and industries. Activities to coordinate accelerator magnet manufacturing could be established in ways that cross boundaries of national laboratories and harvest the wealth of innovations in academia. Simplification of partnership mechanisms between national labs and industry would be helpful, such as via technical services frameworks and updates structures for collaborative research agreements.

- Provide industry access to measurements and test facilities at low cost. Quality plans associated with major procurements usually place the burden of quality control on the conductor supplier. However, significant time and expense is required to set up quality control capabilities, qualify them via benchmarks and inter-laboratory comparisons, and adequately carry out quality measurements. User facilities for industries at national laboratories could be helpful. For HTS conductors like REBCO, the present in-line capabilities often do not probe the conditions for end use of the product. Feedback between quality measurements and processing must be tightly looped, especially during R&D phases.

## *Potential favorable disruptive factors in science and industry*

Close engagement of the HEP magnet development community with other areas of magnet research and development could prove to be very helpful for magnet conductor development.

Privately funded fusion energy research is acquiring REBCO conductor at the scale of a large science facility project, with ~$1 billion overall investment and a significant portion, ~$100 million, landing on conductor. This should drive several positive activities reminiscent of the virtuous cycle of development for Nb-Ti conductors:

- Magnet "pull" is being created by the vision and objectives of the fusion projects. A huge amount of credit is owed to the project leaders and their supporters.
- Multiple vendors are competing to provide conductor that meets a common specification. If the specification and conductor samples can be shared with publicly funded research groups, then a basis of important processing – structure – property links can be determined for each conductor, and research to reduce or eliminate property variations can begin. Coupling to lab and university groups could provide diagnosis and feedback to the manufacturers. Programs like INFUSE are already allowing a start on such collaborations.
- Techniques are being developed to measure $J_c$ properties at the field, field angle, and temperature of anticipated end use, besides the usual manufacturer characterizations at 77 K and self-field. The related measurement science, e.g. that performed for high-field magnet projects such as the 32T user magnet at NHMFL [79], is undergoing continuous activity in response. The high current density of REBCO makes transport measurements especially vulnerable to burn-out, and magnetometry is emerging as an alternative technique [112]. Better networking between the laboratories, universities, and private projects could improve the capability of manufacturers and partner labs to provide quality assurance relevant for end-use conditions.
- Many small test magnets are being manufactured and tested to conductor limits, but there is still much confidentiality associated both with successes and failures of HTS magnets. An important lesson of the Nb-Ti Workshop is that manufacturers originally came to the workshop with a view that they each had an optimized process that was proprietary and special. The strong interactivity of presentation and discussion at the workshops soon lead to entirely new ways of looking at optimization and the ultimate potential of Nb-Ti, leading to the recognition that sharing of experience drove an ultimately much more rewarding technology. If the results of present magnet testing within the various laboratories and research groups could also be shared, broad discussion could occur about potential conductor vulnerabilities exposed by tests. These discussions pull the materials characterizations and help define the topics for PhD research in university groups.

Superconducting rotating machines for sustainable energy are presently entering the marketplace. Many activities worldwide aim to develop wind turbines in the 15–25 MW range [113], [114]. Propulsion for ships, trains, and airplanes is also under development based on the significant increase in power density supplied by high-field superconducting coils. For example, electric aircraft are envisioned to significantly exceed 10 kW / kg and displace high-bypass turbofans as the propulsion unit [115]. Racetrack field coil configurations in many machines have synergy with dipole magnets in accelerators, so there could be opportunity for public-private partnerships. Many envisioned applications operate at temperature higher than 10 K, for which HTS and possibly iron-based superconductors are attractive. A 10 GW wind farm could require procurement of 100 tons of conductor.

Changes in medical diagnosis and therapy guidelines could drive increased development of high-field imaging systems, particle-beam therapy systems, and high-field NMR tools for *in vivo* characterization. Industry has begun to put 30-tesla class NMR systems into service, which require HTS conductors, but results from using these systems is only beginning to disseminate to the medical community. Industry has also begun to deploy compact synchrotron sources using high-field magnets on gantries, but these are in early stages of market penetration and the business drivers are still clarifying. World-class facilities like ISEULT have just begun courses of scientific research relevant for diseases of the brain and other potential therapies where extremely high resolution is needed.

## Appendix 2

*Methodology of assembling, polling, and facilitating stakeholder discussions.*

ARDAP Workshop on the Development of Business Models for Superconductors in the Accelerator Sector Supply Chain

13-14 March 2022

Tufts University

1

## Phase 1 – Interviews prior to workshop

2

## Setting the stage for the meeting - interviews

- Interviewed 21 people from 13 different organizations academic, laboratory and commercial
- Explained the intent of interviews to each, encouraging divergent ideas
- Asked the same questions of each
- Promised anonymity

3

## Findings/general insights from interviews

- Distinction between products and projects
- Distinction between systems and components
- Critical shortages in talent, as well as materials and manufacturing
- Timeline to get out of the business is as short as 18 months for some businesses
- No feasible near term new commercial application
- Market forces are the single most important component to manage SC
- Must provide DOE actionable items

4

## Themes and recommendations from interviews

- Develop a better narrative on why these issues matter to the nation
- Government funded $10M/yr floor is enough to keep capability 'warm'
- Better resource planning - stockpiling, sharing etc
- Improve transparency between national labs and industry
- Gov't, academia and industry hub requires resolving IP issues

5

## Phase 2 – In-person Workshop

6

## Workshop methodology

- Define the collective narrative
- Build “best of breed” descriptions of the challenges and solutions
  - Four perspectives: US Leadership, Innovators, Suppliers, and Buyers
- Advance the narrative toward discussion of business models

7

## In-person attendance:

US Leadership
Buyers
Suppliers
Innovators

| Last Name | First | Affiliation | Email | RSVP |
|---|---|---|---|---|
| Chiesa | Luisa | Tufts University | luisa.chiesa@tufts.edu | host |
| Cooley | Nancy | | lnov8887@gmail.com | host |
| Amm | Kathleen | BNL | ammk@bnl.gov | in-person |
| Babel | Andrew | Raytheon | andrew.babel@raytheon.com | in-person |
| Carl | Matthew | ATI Specialty Metals | matthew.carl@atimetals.com | in-person |
| Cooley | Lance | ASC/NHMFL/FSU | ldcooley@asc.magnet.fsu.edu | in-person |
| Ellis | Tyler | Commonwealth Fusion | tyler@cfs.energy | in-person |
| Hazelton | Drew | SuperPower Inc | Hazelton, Drew <dhazelton@superpower-inc.com> | in-person |
| Jones | Mark | Mevion Medical Systems | Mjones@mevion.com | in-person |
| Kshirsagar | Parag | Raytheon | parag.kshirsagar@rtx.com | in-person |
| Larbalestier | David | ASC/NHMFL/FSU | larbalestier@asc.magnet.fsu.edu | in-person |
| Martin | Matthew | Oxford Instruments Nanoscience | Matt.martin@oxinst.com | in-person |
| Minervini | Joe | MIT | minervini@psfc.mit.edu | in-person |
| Naughton | Miles | ATI Specialty Metals | miles.naughton@atimetals.com | in-person |
| Palmer | Mark | BNL | mpalmer@bnl.gov | in-person |
| Parizh | Michael | GE | parizh@ge.com | in-person |
| Parrell | Jeff | Bruker OST | jeffrey.parrell@bruker.com | in-person |
| Peng | Xuan | HyperTech | xpeng@hypertechresearch.com | in-person |
| Prestemon | Soren | Lawrence Berkeley National Laboratory | soprestemon@lbl.gov | in-person |
| Smathers | David | Materion (formerly HC Starck) | david.smathers@hcstarcksolutions.com | in-person |
| Solovyov | Slowa | Brookhaven Technology Group | slowa@brookhaventech.com | in-person |
| Tentnowski | Michael | FSU | mtentnowski@fsu.edu | in-person |
| Twin | Andrew | Oxford Instruments Nanoscience | Andrew.TWIN@oxinst.com | in-person |
| Vail | Brian | | brian@fbasimplified.com | in-person |
| Hischier | Whitney | UC Berkeley | hischier@berkeley.edu | instructor |
| Rotkoff | Steven | retired | redtmbyrotkoff@icloud.com | instructor |

8

# Day 1

| Time | Activity |
|---|---|
| 9:30-10 am | Intros, background on red teaming, rules of engagement |
| 10-10:15 | Interview findings |
| 10:15-10:45 | 5 will get you 25 |
| 10:45-11 | Break |
| 11:00 - 12:30 pm | Ad Agency |
| 12:30 - 12:45 | Pick up Lunch |
| 12:45-2:15 | Outbriefs of Ad agency: 10 min per team x 6 teams, and collective agreement on group narrative critical points (30 minutes) |
| 2:15-4:45 | Supply chain exercise in groups (can break as needed) |
| 3:00 | Refreshments |
| 4:45 - 5:00 | Conclusion of Day 1 and homework assignments |

9

# Day 2

| Time | Activity |
|---|---|
| 9:00 - 9:30 am | Homework debrief |
| 9:30 - 11:00 | Conclusion of supply chain exercise |
| 11:00 - 11:15 | Break |
| 11:15 - 12:00 pm | Revisit of 5 will get you 25 |
| 12:00 - 12:30 | Lunch |
| 12:30 - 1:00 | Finalize and affirm a collective agreement of workshop outcomes |
| 1:00 - 1:45<br>+ ZOOM | Discussion of workshop outcomes with DOE and external attendees<br>• Interview findings<br>• Outbrief of collective narrative<br>• Outbrief of supply chain exercise |
| 1:45 - 2:00 | Discussion of upcoming actions and deliverables to DOE |
| 2:00 pm | Close of workshop |

10

Red Teaming Team:
Col Steve Rotkoff

Whitney Hischier

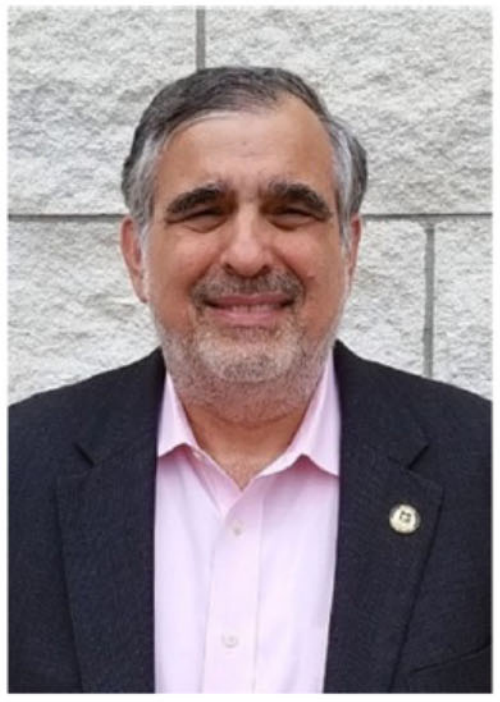

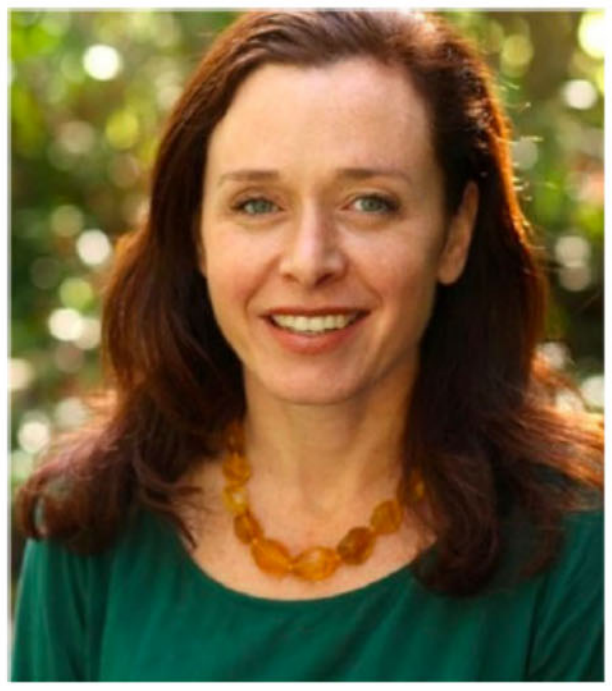

11

## Workshop mission

Scientific developments have created a divergence between the needs of DOE in support of research projects and the superconducting magnet marketplace. DOE requirements are unique, usually without commercial application, and require long time to develop.

As a consequence DOE business is not steady, but episodic and there is no business incentive to invest in material, manufacturing, or talent in areas supporting the superconducting high field magnet ecosystem. The most likely commercial application have a lead time of 5 years or more.

Without a strategy, US commercial capability in this critical technology will atrophy in the next 18 months and wither after that.

***How can we collectively prevent this from happening?***

12

## Rules of engagement

- We start on time, we end on time, we break on time, we lunch on time - please respect the schedule
- Honest, constructive feedback is critical
- Opportunities to speak shall be balanced—please make way

13

## What is red teaming?

Red Teaming is:

a flexible cognitive approach to thinking, planning, and decision making.

Red Teaming uses structured tools and techniques to:

ask better questions
challenge explicit and implicit assumptions
develop alternatives

shift between multiple perspectives to better understand complex environments

to ensure the BEST idea wins, vs that with the strongest advocate

14

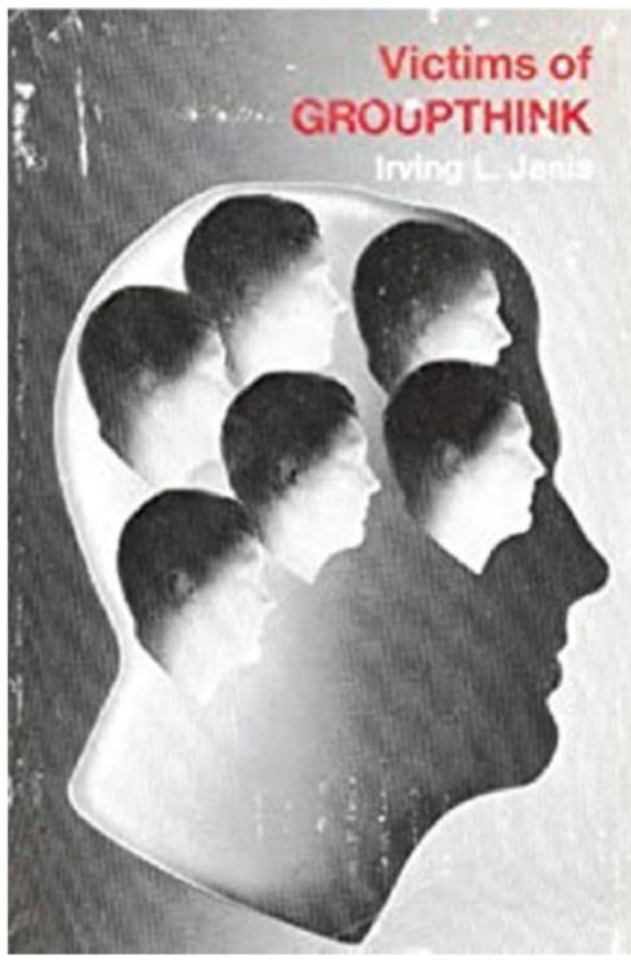

- **Illusions of invulnerability:** the belief that the group cannot fail.
- **Morality:** the belief that the group's motives are inherently good and correct.
- **Rationalizing:** the tendency to "explain away" contradictory information or data.
- **Stereotyping:** the tendency to portray others who are opposed to the group as evil or stupid.
- **Self-censorship:** the tendency of group members to keep their doubts to themselves.
- **Illusions of unanimity:** the belief that silence is the same as agreement.
- **Mindguards:** the emergence of self-appointed thought police who actively shield the group from information that might challenge its assumptions.

**Conformity:** the tendency to view dissent as disloyalty.

15

## Goals for workshop – result from a collective exercise

- Conceive a plan to anchor manufacturing of advanced superconductors in the US
- Develop a paradigm for ARDAP investment in superconductor development and procurement that will resonate with Congress
  - Provide support to the industry – lab – academia ecosystem
- Identify crosslinks between technologies
- Leverage other agencies
- Contemplate product development life-cycles

16

## Goals for workshop – themes

- Keep $Nb_3Sn$ industry and supply chain going
- Diversify network/conversation/perspectives
- Continuity
- Where are future applications to help people live better
- How to make the right decisions
- How to level demand
- How to continue to innovate
- How academic community sees magnets
- Speed to process/manufacture
- Partnering between labs and industry
- Commercializing HTS
- HTS supply chain
- Moving LTS to HTS
- How specific companies fit in overall supply chain
- Define ecosystem to reliably specify future needs and sustain them
- Successful ecosystem

17

## Shortages vs potential solutions

**Shortages**

**Potential Solutions**

| | Raw resources / materials | Talent and training | Equipment and capabilities | Innovation engine to develop new products |
|---|---|---|---|---|
| Stockpiling | | | | |
| Acceleration of commercial applications | | | | |
| Steady state public private partnership | | | | |
| Network adjacent technology areas | | | | |

18

Mitigating supply chain risks in
# Equipment and Facilities

19

## Equipment and capabilities

- Near term:

| Problem | Countermeasures |
|---|---|
| Lack of US industrial capability for accelerator magnet manufacture | - ARDAP Incubator (TRL-graded $)<br>- Stewardship program (enhanced)<br>- Tech transfer<br>- Engage private industry with opportunities |
| Poor business case justification for sustaining / renewing major capitol equipment and facilities | - Stockpile minimum annual purchases<br>- Labs then take from stockpile<br>- Stimulate / incentivize ecosystem collaboration Gov/Acad./Industry working towards a common, defined goal |

20

## Equipment and capabilities

- Mid term:

| Problem | Countermeasures |
|---|---|
| Small scale trial production of S/C is not economical | Maintain "warm" production through regular procurements of state of the art conductor<br>=> Creates "room" for trials at small-scale with quality benefits of "real" production |
| Need greater access / availability of:<br>- large bore test facility<br>- Cable test facility<br>- High. Field (>20T) measurement systems | -HTS Cable Test Facility at FNAL<br>- Additional >20T systems (solenoids) for general purpose use (results could demonstrate conductor improvements) |
| Need more experience / trials with HEP-scale Nb3Sn magnet manufacturing | Make conductor more available via "stockpile" (see: annual purchase of conductor above) |

21

## Mitigating supply chain risks in Raw Materials

22

## Challenges, insights, and findings

- Key shortages and threats in many areas
- Making a material to a specification requires a persistent base load of activity to ensure quality and delivery
- Instabilities and uncertainties on global scale

23

## How to leverage the ecosystem to achieve your goals

- Raw Materials Supply
  - Precious Materials Examples: Helium, Nb and Rare Earths
    - Maintaining a stable supply is critical for superconducting technology development
    - For Helium, recovery and recycling needs to be encouraged
    - Need to understand the full range of potential users
    - This argues for a stockpile approach to ensure availability of critical materials
      - However, how do we implement this in a way such that government isn't competing with industry for day-to-day supply of the commodity
      - Who will have access to the stockpile?
        - Is this programmatic?
        - Is this emergency situation access (i.e. strategic reserve)?
        - Is this dedicated to R&D?
          - Who's R&D – agency-funded vs industrial
      - Are there economies of scale for bulk purchases?
  - Bulk Commodity Materials Examples: Copper
    - Stockpiling isn't reasonable for commodity materials
    - Encourage development of domestic supply
      - Perhaps as a capability that could be re-started
      - Warm capability – is there a government funding mechanism for this?
    - Identify potential domestic supply points
    - Explore mechanisms to reduce environmental impacts

24

## How to leverage (part 2)

- HTS conductor
  - Can we develop a predictive model of supply and demand?
  - Limited stockpiles may work
    - Provide short term supplies
  - Major stockpiles are undesirable
    - Development is still underway
    - Specifications are rapidly evolving
  - Applications development can't move forward without sufficient short-term supplies

25

## Opportunities and roadblocks to PPPs

- Intellectual property remains the biggest hurdle
  - The legal teams need to be included in the early-stage planning and discussions
- PPP might enable new mechanisms to stockpile critical materials for development needs

26

Mitigating supply chain risks in

# Innovation and stimulation of ideas

27

## Challenges, findings, and insights

- Lack of support through the entire product development cycle
- Shoulder manufacturing next to production lines
- Innovation of new ideas needs to connect to developmental innovation and manufacturing alternatives
- Opportunities are presented by related activities
- Use of a conductor requires more than the conductor – ancillary technology

28

## Goals for innovation and stimulation of ideas

- Sustained innovation incentives above / beyond SBIR and the present status of CPRD, Early Career Awards, LDRD, and University innovation
- Facilitated bridging from innovation to specification-driven commercialization
- Dedicated facilities and know-how for scaling innovations from initiation through development to production
- Manufacturing alternatives and "design for manufacturing" (D4M)
- Grow capability in measurement science and instrumentation, especially new knowledge accumulation approaches
- Ensure simultaneous innovation and advancement of ancillary technologies

29

## How to leverage the ecosystem to achieve your goals

1. Make steady (annual) sizeable procurements of advanced conductors and stock a national conductor repository useable by research and industry
2. Establish a magnet innovation institute or technology hub including manufacturing infrastructure for development and scaling
3. Fund and build technology demonstrations
4. Invest public funds to support component development, e.g. the magnet that goes into a system
5. Support a flexible national network to develop / respond to promising applications

30

## Opportunities and roadblocks to PPPs

- Entities and activities exist that could be effectively networked by ARDAP funds:
  - MDP, CPRD etc in HEP; INFUSE providers;
  - A revitalized industry consortium?
- Infrastructure exists that could be included in the above network
- The academic – lab – industry nexus grown by HEP can be used as a model for virtuous development cycles

31

## Mitigating supply chain risks in
## Talent and Training

32

## Findings and comments

- Industry builds teams to meet product goals
  - “Warm” production for the future accelerator sector is a different configuration than that for profitable production of e.g. MRI wire
- “Warm” manufacturing requires investment at the highest level of base load to sustain capability

33

## How to leverage the ecosystem to achieve your goals

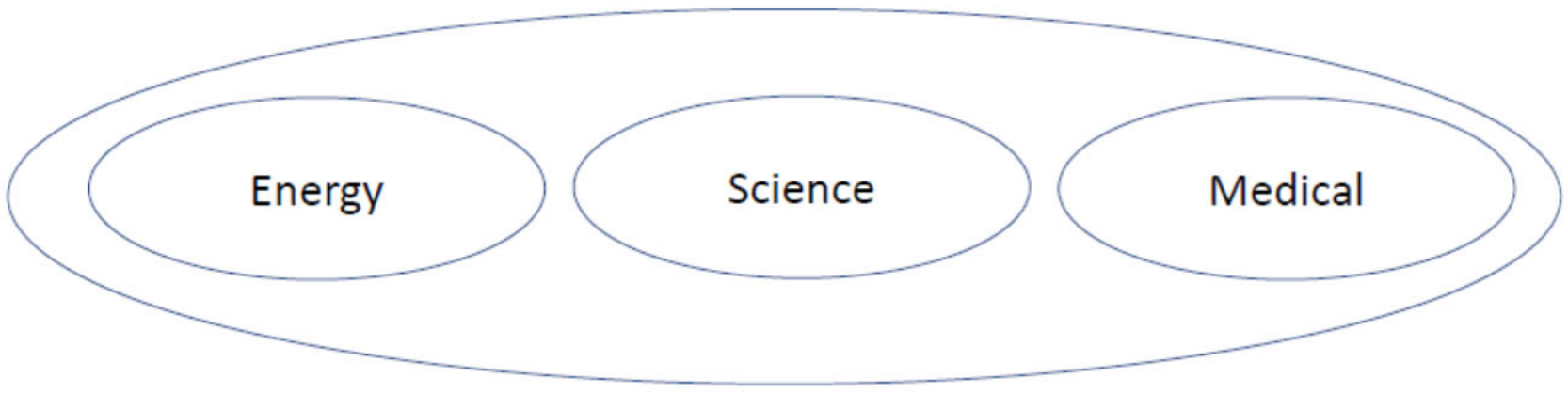

| | | | | |
|---|---|---|---|---|
| Past | Generator, T&D demonstrators | Tevatron, LHC, Semiconductor materials development | MRI, NMR | |
| Future | Offshore wind, Compact Fusion, Aviation, T&D | EIC, FCC, ILC, ITER, Muon, Quantum computing | Proton therapy, Advanced Neuroimaging | MASA |

34

## How to leverage the ecosystem to achieve your goals

- Leverage grass roots efforts – (College career engagement plans, ASEF, IEEE CSC, Accelerator Stewardship) – coordinate
- Make the funding visible – important for seeing the future
- Create management cultures that encourage career development, knowledge management
- Get national labs to share career frameworks. Look at national models
  - A career framework is **a blueprint of the requisite roles, talent capabilities and accountabilities for supporting execution of an organization's business strategy**
- Along with national career frameworks – look at training in technical leadership/development at all levels

35

## Opportunities and roadblocks to PPPs

- Make funding contingent on creating the attractive narrative, environment and culture to make Superconductivity a high demand field
- Stove piping is a challenge for knowledge management and transfer due to IP, national security elements, knowledge hording
- Seek to align career frameworks between industry and government career frameworks to allow career mobility

36

## Output from workshop – next steps

Address the following questions articulated by DOE:

- What is a feasible business approach to develop or enhance domestic capabilities to meet the needs of science facilities in the superconducting accelerator systems sector?
- What is the optimal public / private partnership relative to the business approach above? How should key functions be carried out, including:
    - Technology development, transfer, and management of intellectual property
    - Management and allocation of public funding
    - Interactions among laboratories, universities, and industry
    - Testing, measurements, and assurance of quality
- What are sustainable pathways to markets at scale?
- What are needs for workforce development and training?

37